\documentclass[longauth]{aa}  
\usepackage{graphicx}
\usepackage{natbib}
\usepackage{txfonts}
\usepackage[colorlinks=true, citecolor=blue]{hyperref}
\usepackage{multirow}
\usepackage{adjustbox}
\usepackage{comment}
\usepackage{caption}
\usepackage{afterpage}
\usepackage{orcidlink}

\begin{document} 
   \title{The ALMA survey to Resolve exoKuiper belt Substructures (ARKS)} \subtitle{IV: CO gas imaging and overview}
\author{S. Mac Manamon\textsuperscript{1}\orcidlink{0000-0002-7066-8052}\fnmsep\thanks{E-mail: macmanas@tcd.ie} 
\and L. Matr\`a\textsuperscript{1} \orcidlink{0000-0003-4705-3188}
\and S. Marino\textsuperscript{2} \orcidlink{0000-0002-5352-2924}
\and A. Brennan\textsuperscript{1} \orcidlink{0000-0002-7050-0161}
\and Y. Han\textsuperscript{3} \orcidlink{0000-0002-2106-0403}
\and M. R. Jankovic\textsuperscript{4} \orcidlink{0000-0001-6684-6269}
\and P. Weber\textsuperscript{5,6,7}\orcidlink{0000-0002-3354-6654}
\and M. Bonduelle\textsuperscript{8} \orcidlink{0009-0000-7049-8439}
\and J. M. Carpenter\textsuperscript{9} \orcidlink{0000-0003-2251-0602}
\and G. Cataldi\textsuperscript{10,11} \orcidlink{
0000-0002-2700-9676}
\and A. M. Hughes\textsuperscript{12} \orcidlink{0000-0002-4803-6200}
\and A. K\'osp\'al\textsuperscript{13,14,15}\orcidlink{0000-0001-7157-6275}
\and J. P. Marshall\textsuperscript{16} \orcidlink{0000-0001-6208-1801}
\and B. C. Matthews\textsuperscript{17,18} 
\and J. Milli\textsuperscript{8} \orcidlink{0000-0001-9325-2511}
\and A. Mo\'or\textsuperscript{13} \orcidlink{0009-0001-9360-2670}
\and K. \"Oberg\textsuperscript{19} \orcidlink{0000-0001-8798-1347}
\and S. P\'erez\textsuperscript{5,6,7} \orcidlink{0000-0003-2953-755X}
\and A. A. Sefilian\textsuperscript{20} \orcidlink{0000-0003-4623-1165}
\and D. J. Wilner\textsuperscript{19} \orcidlink{0000-0003-1526-7587}
\and M. C. Wyatt\textsuperscript{21} \orcidlink{0000-0001-9064-5598}
\and E. Chiang\textsuperscript{22} \orcidlink{0000-0002-6246-2310}
\and A. S. Hales\textsuperscript{23,9,6} \orcidlink{0000-0001-5073-2849}
\and J. B. Lovell\textsuperscript{19} \orcidlink{0000-0002-4248-5443}
\and P. Luppe\textsuperscript{1} \orcidlink{
0000-0002-1018-6203}
\and M. A. MacGregor\textsuperscript{24} \orcidlink{0000-0001-7891-8143}
\and T. Pearce\textsuperscript{25} \orcidlink{0000-0001-5653-5635}
\and M. Booth\textsuperscript{26} \orcidlink{0000-0001-8568-6336}
\and C. del Burgo\textsuperscript{27,28} \orcidlink{0000-0002-8949-5200}
\and A. Fehr\textsuperscript{19}\orcidlink{0000-0001-7155-3583}
\and E. Mansell\textsuperscript{12}\orcidlink{0009-0004-1433-8149}
\and B. Zawadzki\textsuperscript{12} \orcidlink{0000-0001-9319-1296}}

\institute{
School of Physics, Trinity College Dublin, the University of Dublin, College Green, Dublin 2, Ireland \and
Department of Physics and Astronomy, University of Exeter, Stocker Road, Exeter EX4 4QL, UK \and
Division of Geological and Planetary Sciences, California Institute of Technology, 1200 E. California Blvd., Pasadena, CA 91125, USA \and
Institute of Physics Belgrade, University of Belgrade, Pregrevica 118, 11080 Belgrade, Serbia \and
Departamento de Física, Universidad de Santiago de Chile, Av. V\'ictor Jara 3493, Santiago, Chile \and
Millennium Nucleus on Young Exoplanets and their Moons (YEMS), Chile \and
Center for Interdisciplinary Research in Astrophysics Space Exploration (CIRAS), Universidad de Santiago, Chile \and
Univ. Grenoble Alpes, CNRS, IPAG, F-38000 Grenoble, France \and
Joint ALMA Observatory, Avenida Alonso de C\'ordova 3107, Vitacura 7630355, Santiago, Chile \and
National Astronomical Observatory of Japan, Osawa 2-21-1, Mitaka, Tokyo 181-8588, Japan \and
Department of Astronomy, Graduate School of Science, The University of Tokyo, Tokyo 113-0033, Japan \and
Department of Astronomy, Van Vleck Observatory, Wesleyan University, 96 Foss Hill Dr., Middletown, CT, 06459, USA \and
Konkoly Observatory, HUN-REN Research Centre for Astronomy and Earth Sciences, MTA Centre of Excellence, Konkoly-Thege Mikl\'os \'ut 15-17, 1121 Budapest, Hungary \and
Institute of Physics and Astronomy, ELTE E\"otv\"os Lor\'and University, P\'azm\'any P\'eter s\'et\'any 1/A, 1117 Budapest, Hungary \and
Max-Planck-Insitut f\"ur Astronomie, K\"onigstuhl 17, 69117 Heidelberg, Germany \and
Academia Sinica Institute of Astronomy and Astrophysics, 11F of AS/NTU Astronomy-Mathematics Building, No.1, Sect. 4, Roosevelt Rd, Taipei 106319, Taiwan. \and
Herzberg Astronomy \& Astrophysics, National Research Council of Canada, 5071 West Saanich Road, Victoria, BC, V9E 2E9, Canada \and
Department of Physics \& Astronomy, University of Victoria, 3800 Finnerty Rd, Victoria, BC V8P 5C2, Canada \and
Center for Astrophysics | Harvard \& Smithsonian, 60 Garden St, Cambridge, MA 02138, USA \and
Department of Astronomy and Steward Observatory, The University of Arizona, 933 North Cherry Ave, Tucson, AZ, 85721, USA \and
Institute of Astronomy, University of Cambridge, Madingley Road, Cambridge CB3 0HA, UK \and
Department of Astronomy, University of California, Berkeley, Berkeley, CA 94720-3411, USA \and
National Radio Astronomy Observatory, 520 Edgemont Road, Charlottesville, VA 22903-2475, United States of America \and
Department of Physics and Astronomy, Johns Hopkins University, 3400 N Charles Street, Baltimore, MD 21218, USA \and
Department of Physics, University of Warwick, Gibbet Hill Road, Coventry CV4 7AL, UK \and
UK Astronomy Technology Centre, Royal Observatory Edinburgh, Blackford Hill, Edinburgh EH9 3HJ, UK \and
Instituto de Astrof\'isica de Canarias, Vía L\'actea S/N, La Laguna, E-38200, Tenerife, Spain \and
Departamento de Astrof\'isica, Universidad de La Laguna, La Laguna, E-38200, Tenerife, Spain
}
   \date{Received 25 July 2025; accepted 24 September 2025}

  \abstract 
   {CO gas is detected in a significant number ($\sim 20$) of debris discs (exoKuiper belts), but understanding its origin and evolution remains elusive. Crucial pieces of evidence are its mass and spectro-spatial distribution, which are coupled through optical depth and have only been analysed at low to moderate resolution so far. The ALMA survey to Resolve exoKuiper belt Substructures (ARKS) is the first ALMA large program to target debris discs at high spectro-spatial resolution. }
   {We used $^{12}$CO and $^{13}$CO J=3-2 line data of 18 debris belts observed by ARKS, 5 of which were already known to be gas-bearing, in order to analyse the spectro-spatial distribution of CO and constrain the gas mass in discs that were known to host gas previously, and to search for gas in the remaining 13 discs without previous CO detections.}
   {We developed a line-imaging pipeline for ARKS CO data with a high spectro-spatial resolution. Using this tool, we produced line cubes for each of the ARKS targets, with a spatial resolution down to about 70 mas and a spectral resolution of 26 m~s$^{-1}$. We used spectro-spatial shifting and stacking techniques to produce a gallery of maps with the highest possible signal-to-noise ratio (S/N) and with radial and spectral profiles that reveal the distribution and kinematics of gas in five gas-bearing discs at unprecedented detail. } 
   {For each of the five gas-bearing discs (HD~9672/49~Ceti, HD~32297, HD~121617, HD~131488, and HD~131835), we constrained the inner radius of the $^{12}$CO ($r_{\rm min} \sim 3 - 68$ au), and we found that the radial brightness profile of CO peaked interior to the dust ring, but that CO was also more radially extended than the dust. In a second-generation scenario, this would require significant shielding of CO that would allow it to viscously spread to the observed widths. We present the first radially resolved $^{12}$CO/$^{13}$CO isotopologue flux ratios in five gas-bearing debris discs and found them to be constant with radius for the majority (four out of five) of systems. This indicates that $^{12}$CO and $^{13}$CO are both optically thick or optically thin throughout the discs. We report CO line fluxes or upper limits for all systems and optical depth dependant masses for the five systems with detected CO. Finally, we analysed the $^{12}$CO J=3-2 line luminosities for a range of ARKS debris discs and for debris discs from the literature. We confirm that gas is mostly detected in young systems. However, the high scatter seen in young/high fractional luminosity systems indicates no trend within the systems with detected gas. This could be caused by different system properties and/or evolution pathways. }
   {}
   \maketitle
   
\section{Introduction} 
Planetesimal belts (debris discs) are belts of icy dust, comets, asteroids, and planetesimals. They are analogous to the Kuiper and asteroid belts in the Solar System \citep[for reviews, see][and references therein]{wyatt2008evolution, matthews2014observations, hughes2018debris}. These belts can be observed because the surface area that is covered by small dust grains is so large. The dust, comets, asteroids, and planetesimals collide constantly with one another, grind each other down, and create more dust. Debris discs were originally thought to be gas-free environments that solely consist of solids, but molecular gas emission has recently been observed in multiple discs \citep[e.g.][]{kospal2013alma, moor2013alma, moor2017, dent2014molecular, lieman2016debris, matra2015co_fomalhaut, matra2017exocometary_betapic}. We now know of over 20 debris discs (most commonly found around intermediate-mass stars) that contain gas \citep{marino2020population}, and we know this not only through molecular emission observed at millimetre (mm) wavelengths, but also atomic C and O emission in the far-infrared (IR) \citep[e.g.][]{cataldi2014herschel, brandeker2016herschel, roberge2013herschel}, CO absorption in the near-IR \citep{troutman2011ro}, and CO and atomic absorption in the ultraviolet (UV) \citep[e.g.][]{roberge2000high, roberge2014volatile, wilson2019detection,Wu2024_argon_betapic, brennan2024low}. CO is the most commonly observed species (and the only molecule) to be observed in debris discs, however, which is largely due to the significant increase in sensitivity provided by the Atacama Large Millimetter/submillimeter Array (ALMA) \citep[e.g.][]{matra2019ubiquity_TWA7, klusmeyer2021deep, smirnov2022lack}. 

Two possible origins have been proposed for the debris disc gas. The first proposition was that the gas is of primordial origin and is left over from the protoplanetary disc phase. This would be surprising because gas is expected to have dissipated by the time the debris disc formed. Gas dissipation is thought to be a pre-condition for the escalation of collisional evolution of the planetesimals that characterise a debris disc \citep{wyatt2008evolution}. The second proposition was that the gas is of secondary origin and was released from the planetesimals via collisional evolution of a shared origin with the dust.

In the primordial scenario, the gas could potentially remain in discs around A-type stars due to inefficient photoevaporation of protoplanetary discs, around intermediate-mass stars if their discs were depleted in small grains ($\lesssim 0.01$  $\mu$m) so that far ultraviolet (FUV) photoelectric heating is ineffective, thus prolonging the disc lifetime through a delayed photoevoporation of the disc \citep{nakatani2021photoevaporation, nakatani2023primordial}, and in general, if the gas densities are high enough that the photodissociation lifetime of CO gas can be extended through shielding \citep{kospal2013alma}. Without any shielding, the lifetime of a CO molecule against photodissociation by the interstellar radiation field would be 130 years \citep{heays2017photodissociation} because there is not enough dust to shield the CO from the stellar and the interstellar UV radiation field \citep{kospal2013alma}.

The CO masses in CO-rich debris discs \citep[e.g. HD21997, HD~121617, HD~131488, HD~131835;][]{kospal2013alma, moor2017, hales2019modeling} were inferred to be sufficiently high that if H$_2$, the dominant shielding agent for CO in a primordial scenario, were present in primordial amounts (with a primordial CO/H$_2$ ratio), it could allow the CO to survive for the entire disc lifetime. The origin of CO in these systems could therefore plausibly be primordial. \cite{kospal2013alma} found that for these young debris discs (10-20 Myr) with masses of CO that are too high to be explained by just the destruction of exocomets, gas that persisted from the protoplanetary disc phase is a likely explanation. In particular, the masses of CO in these discs are of the order of $M_\mathrm{CO} \gtrsim 10^{-3} M_\oplus$ \citep{cataldi2023primordial}.

As mentioned above, the same processes that create dust in debris discs are also capable of releasing gas and are a viable alternative explanation for the origin of gas in debris discs. Gas can be released by sublimation of icy planetesimals (exocomets), photodesorption from dust grains \citep{grigorieva2007survival}, vaporisation of colliding dust particles \citep{czechowski2007collisional}, or collision of exocomets/icy planetesimals \citep{zuckerman201240, kospal2013alma, bonsor2023secondary}. In this second-generation gas scenario, which is based on the composition of comets in the Solar System, the gas is dominated by CO, CO$_2$, and H$_2$O, with much smaller amounts of H$_2$ compared to a primordial scenario \citep[e.g.][]{kospal2013alma}. CO self-shielding and shielding by C I, a photodissociation product of CO, are likely to play a significant role in extending the lifetime of the gas in the more CO-rich discs if they are secondary in origin \citep[e.g.][]{matra2017exocometary_betapic, kral2019imaging, marino2020population}. 

Gas-removal processes operate continuously in debris discs. CO can be depleted via photodissociation and freeze-out onto dust grains, though the latter is inefficient at the temperatures and small collective surface area of dust grains in debris discs \citep{matra2015co_fomalhaut}. For gas to be observable in debris discs, it must therefore be continuously replenished and/or shielded from photodissociation over lifetimes comparable to the system age \citep{kospal2013alma}.

For low-mass CO (or CO-poor) discs (e.g. $\beta$ Pic, Fomalhaut, HD~181327, TWA 7, and $\eta $  Corvi), the survival timescale of CO is much shorter than the ages of their systems, even if the CO is being shielded by unseen H$_2$ \citep[][]{dent2014molecular, matra2018molecular_betapic, matra2017detection_fomalhaut, marino2016exocometary,marino2016alma}. Gas production by exocomets can explain discs with low CO masses ($M_\mathrm{CO} \lesssim 10^{-4} M_\oplus$) because the required production rate of CO agrees reasonably well with the destruction rates of exocomets in these debris discs \citep{cataldi2023primordial}. 

If the gas is being replenished from solids (icy dust or larger exocomets), as in the secondary origin scenario, the naive expectation would be that the gas should be radially co-located with the dust component, as was found by \cite{dent2014molecular} and \cite{matra2017exocometary_betapic} for $\beta$ Pic. However,  we expect the gas to radially viscously spread over time, and to create an accretion disc inwards and a decretion disc outwards, where the viscous spreading rate would depend on the viscosity of the gas \citep{kral2016self}. This was not observed for $\beta$ Pic, however, where the radial distribution of long-lived  C\textsc{I} gas, a daughter product of CO and CO$_2$ destruction, was inconsistent with the expectation of an accretion disc \citep{cataldi2018alma}. If CI production started only $\sim 5000$ yr ago, however, not enough time has passed yet for the gas to viscously spread in to the star to form the accretion disc. \cite{kral2019imaging} similarly studied the radial distributions of C\textsc{I} and CO in the lower-resolution observations of HD~131835 and reported that they were consistent with one another, but inconsistent with an accretion disc. A potential explanation for why an accretion disc might not be visible is that the gas is accreted onto a planet orbiting interior to the disc. 

We investigate the distribution of $^{12}$CO and $^{13}$CO in debris discs at an unprecedented spectro-spatial resolution by exploiting the ALMA survey to Resolve exoKuiper belt Substructure (ARKS, project number 2022.1.00338.L, PI: S. Marino), which is the first ALMA large program to target debris discs. ARKS targets 18 nearby debris discs (and 6 archival sources). The ARKS goals include understanding the diversity and ubiquity of radial and vertical substructures in dust continuum emission and imaging CO in gas-bearing discs at a level that allows us to carry out the first radial, vertical, and gas kinematic study. High-resolution CO observations will allow us to carry out detailed comparisons between the dust and the gas and to constrain the origin and evolution of CO gas in these discs (for sample characteristics and stellar parameters we used, see \citealp{SebasARKSpaper}). 

We used ALMA to observe the $^{12}$CO J=3-2 line (henceforth $^{12}$CO, unless specified otherwise) and $^{13}$CO J=3-2 line (henceforth $^{13}$CO, unless specified otherwise) in order to characterise the spectro-spatial structure of the five previously known gas-bearing targets and conduct a search for $^{12}$CO in the other 13 discs that are as yet devoid of CO gas detections (henceforth gas-free discs). We did not reanalyse the six archival targets, including the CO detection around $\beta$ Pic \citep{matra2019kuiper}, that are included in the ARKS sample because the spectro-spatial resolution is very different and the CO line that was targeted by these archival observations was different as well. In Sect. \ref{observation_sec} we describe the observations and the imaging pipeline we developed for ARKS.  In Sect. \ref{results_sec} we present our results and and discuss their implications in Sect. \ref{discussion_sec}. We conclude with a summary in Sect. \ref{conclusions_sec}.

\section{Observations}
\label{observation_sec}

To image CO gas in the ARKS data, we developed a line-imaging pipeline in Python using the Common Astronomy Software Applications (CASA) package, version 6.4.1.12 \citep{The_CASA_Team_2022}. The detailed calibration and preparation of the interferometric visibilities for imaging are described in \citet{SebasARKSpaper}, but we provide a brief summary here. The calibration of the interferometric visibilities for all observations in time, baseline, and frequency was carried out using the ALMA pipeline. The frequencies were transformed into the barycentric reference frame, and the data were time-averaged to reduce the data size without compromising imaging quality. A phase centre offset (derived from basic fitting of a disc to the continuum visibilities tracing dust emission) was then introduced to ensure no relative astrometric offsets were present across different observing dates for the same system, and to ensure the visibility weights were correct in an absolute sense for each observing date, configuration, and system. Finally, for each of the ARKS sources and each array configuration, two CASA Measurement Sets containing the calibrated visibilities ready for imaging were produced for the spectral windows containing the $^{12}$CO line (rest frequency of 345.796 GHz) and the $^{13}$CO line (rest frequency of 330.599 GHz). 

The $^{12}$CO and $^{13}$CO lines were observed for every system, though the spectral resolution varied depending on whether the system was previously known to be gas-bearing or not. For the 5 gas-bearing discs, the spectral resolution was set to 26 m~s$^{-1}$ (the finest spectral resolution possible with ALMA) for the $^{12}$CO spectral window and to 886 m~s$^{-1}$ for the $^{13}$CO spectral window, which corresponds to two times the native channel width. The gas-bearing discs' $^{12}$CO spectral resolution was set such that the maximum possible kinematic information could be extracted. The $^{13}$CO spectral window uses the maximum bandwidth to not compromise continuum sensitivity. For the discs without previous gas detections, the spectral resolution was set to 846 m~s$^{-1}$ for $^{12}$CO and 886 m~s$^{-1}$ for $^{13}$CO to maximise bandwidth and therefore sensitivity to dust continuum emission. 

For each source and each line, we simultaneously imaged the calibrated and preprocessed visibilities from all baseline configurations using the CLEAN algorithm \citep{hogbom1974aperture}, as implemented through CASA's \verb|tclean| task. We developed an imaging pipeline where \verb|tclean| is called three times. For all three \verb|tclean| iterations implemented as part of our pipeline, the \verb|start|, \verb|width|, and \verb|nchan| parameters represent the barycentric velocity of the first channel, the width of each channel, and the number of channels in the cube outputted by the pipeline. These were set depending on the system and in particular on whether they were known to contain gas; the spectral information used to derive these values was obtained directly from each observations’ Measurement Set header file. The parameter \verb|width| was set to be the same as the observations native channel width, to maximise spectral resolution. In the $^{12}$CO observations, this resulted in channel widths of 13 m s$^{-1}$, which corresponds to a resolution of 26 m s$^{-1}$; in the $^{13}$CO observations, the native channel spacing was 443 m s$^{-1}$, which corresponds to a resolution of 886 m s$^{-1}$. The parameters \verb|start| and \verb|nchan| were calculated based on the expected range of Keplerian orbital velocities of the gas in each system. The range of velocities included in the cube for gas-bearing systems was estimated using the maximum expected Keplerian line-of-sight velocity,

\begin{equation}
\label{vlosmax}
    v_{\rm los, max}= C  v_{\text{los, kep}}(r_{\text{min}})   ,
\end{equation}
where 
\begin{equation}
v_{\text{los, kep}}(r_{\text{min}}) = 30 {\text{ km~s}^{-1}} \sqrt{ \frac{M_\star}{r_{\text{min}}}} \sin{i}
\end{equation}

is the Keplerian projected line of sight velocity at the inner edge $r_{\text{min}}$ (in au) of the disc (given by Eq. \ref{rminmax}), $M_\star$ is the mass of the star in Solar masses, $M_\odot$, estimated as described in Appendix A of \citet{SebasARKSpaper}, and $i$ is the inclination of the disc, measured from the best fit parameters derived by MCMC sampling carried out by \citet[][see Table \ref{tab:system_parameters}]{SebasARKSpaper}. Eq. \ref{vlosmax} corresponds to the maximum line-of-sight velocity at which we expect to detect the gas, given prior ARKS continuum observations of $r_{\rm min}$. In Eq. \ref{vlosmax}, $C$ is a constant factor adjusted for each system following the first iteration of the pipeline to correct for the inaccurate prior on $r_{\rm min}$ and for any departure of gas speed from purely Keplerian velocities (pressure gradients providing radial support), if any, and ensure all emission was included in the cube. Typically, we find values of $C$ between 1.6 and 8.3, depending on the gas-bearing system, reflecting CO disc inner radii that were generally lower than initially estimated based on the dust distribution. For gas-free systems, $v_{\rm los, max}$ was set to 50~km~s$^{-1}$. $v_{\rm los, max}$ was then divided by the width of the channels to calculate \verb|nchan| for the \verb|tclean| calls. \verb|start| was calculated through  $v_{\rm start} = V_* \pm \frac{v_{\rm los, max}}{2}$, where $V_*$ is the stellar velocity in the barycentric frame as obtained from the literature (Table \ref{tab:system_parameters}). The $\pm$ sign depends on whether the visibility file had channels in order of ascending or descending velocity. 

The parameter \verb|restfreq| in each of the \verb|tclean| calls is set to $^{12}$CO and $^{13}$CO rest frequencies. For the gas-bearing discs viewed at an inclination away from edge-on ($i=90^{\circ}$), we employed Briggs weighting, with a robust parameter of 0.5. This gives a good compromise between resolution and sensitivity for these gas-bearing discs. For the close to edge-on gas-bearing discs around HD~32297 and HD~131488, where the emission was spread over a much larger number of channels (leading to lower S/N per beam, per channel), and for the gas-free/non-detected discs, we employed Briggs weighting with a robust parameter of 2, which is close to Natural weighting \citep{briggs1995high}, to achieve maximum sensitivity. 

The first iteration of \verb|tclean| was run without any deconvolution, with a very small pixel size to create a dirty image and obtain the size of the synthesized beam. We used this beam size to optimise the pixel size in later \verb|tclean| runs, which we chose to be $\frac{1}{5}$ of the beam size (averaged between the major and minor axis). We then chose the image size to be equal to the full width at half maximum (FWHM) of the primary beam of the ALMA 12~m antenna. We carried out no deconvolution in the first two rounds of \verb|tclean|. The second iteration of \verb|tclean| is run to create another dirty image, now with an appropriate pixel and image size, which is needed to create a CLEAN mask around the emission in each channel in the cube.

Given the large number of 13 m~s$^{-1}$ channels covering the velocity extent of the $^{12}$CO line ($\gtrsim 1000$), manual masking is unfeasible and masks covering the gas emission in every channel have to be defined automatically. We use a Keplerian mask to leverage the expected near-Keplerian rotation pattern to produce channel-based masks \citep{rich_teague_2020_4321137, czekala2021molecules}. To create these, we use prior knowledge of each disc's geometry, assuming gas is co-located with the dust. Creation of these Keplerian masks is described in full in Appendix \ref{kep_mask}.

In the third and final \verb|tclean| iteration, the deconvolver parameter was set to \verb|multiscale| \citep{cornwell2008multiscale}. We set the scales for the tclean deconvolution to be $[0,5,10]$, corresponding to a point source, a scale of the same size as the synthesised beam, and twice the synthesised beam. The iterative CLEAN process is repeated until the peak emission within all masks in the residual images obtained after subtracting the latest iteration's CLEAN model was below a threshold of 4 times the RMS noise level. We adopted this threshold of 4 times the RMS noise level, in line with the CLEANing methodology of \citet{czekala2021molecules}, as it ensured no significant disc emission remained within the mask in each channel while ensuring no artefacts of the dirty beam was present in the final, CLEANed image, which we confirmed by visual inspection. As is standard in the CLEAN process, the final CLEAN model cube was convolved with the CLEAN beam and added to the final residual cube to obtain final image cubes of $^{12}$CO and $^{13}$CO emission for every disc. Along with these, we produced \verb|FITS| files of the final cubes of the primary beam, the mask used, and the residuals\footnote{
We make these available to the community at \href{https://doi.org/10.7910/DVN/PXGNNZ}{https://doi.org/10.7910/DVN/PXGNNZ} and the imaging pipelines can be found on \url{https://arkslp.org}.}

For the five discs with previously detected gas (HD~9672, HD~32297, HD~121617, HD~131488, HD~131835), the imaging pipeline led to beam sizes ranging from 70 to 150 mas (corresponding to physical scales projected on the sky of 8 to 16 au) for $^{12}$CO and 70 to 160 mas (corresponding to physical scales projected on the sky of 9 to 18 au) for $^{13}$CO. The RMS levels in the $^{12}$CO cubes ranged between 3 to 5 mJy/beam and from 0.6 to 1.0 mJy/beam in the $^{13}$CO cubes (Table \ref{tab:beamsizes_gasrich}).

For the 13 discs not previously known to host CO gas, we only ran the first two \verb|tclean| rounds without masking or deconvolution. This led to beam sizes between 130 to 950 mas (corresponding to physical scales projected on the sky of 9 to 62 au) for $^{12}$CO and 140 to 1000 mas (corresponding to physical scales projected on the sky of 10 to 65 au) for $^{13}$CO. The noise levels were in the range of 0.4 to 1.6 mJy/beam in the $^{12}$CO cubes and 0.5 to 1.6 mJy/beam in the $^{13}$CO cubes (Table \ref{tab:beamsizes_gasrich}).

While we do not include $\beta$ Pic in the analysis in this paper as the archival data covers a different CO line and has a different spectro-spatial resolution to the ARKS targets, the archival data included in the ARKS sample of this disc was imaged in the same manner described here (with Briggs robust parameter set to 0.5) in order to be included in other ARKS papers  \citep[see][]{SebasARKSpaper, Julien_arks}.

\section{Results} 
\label{results_sec}
\subsection{Image galleries}

\begin{figure*}[!ht]
    \centering
    \includegraphics[width = 0.98\textwidth, keepaspectratio]{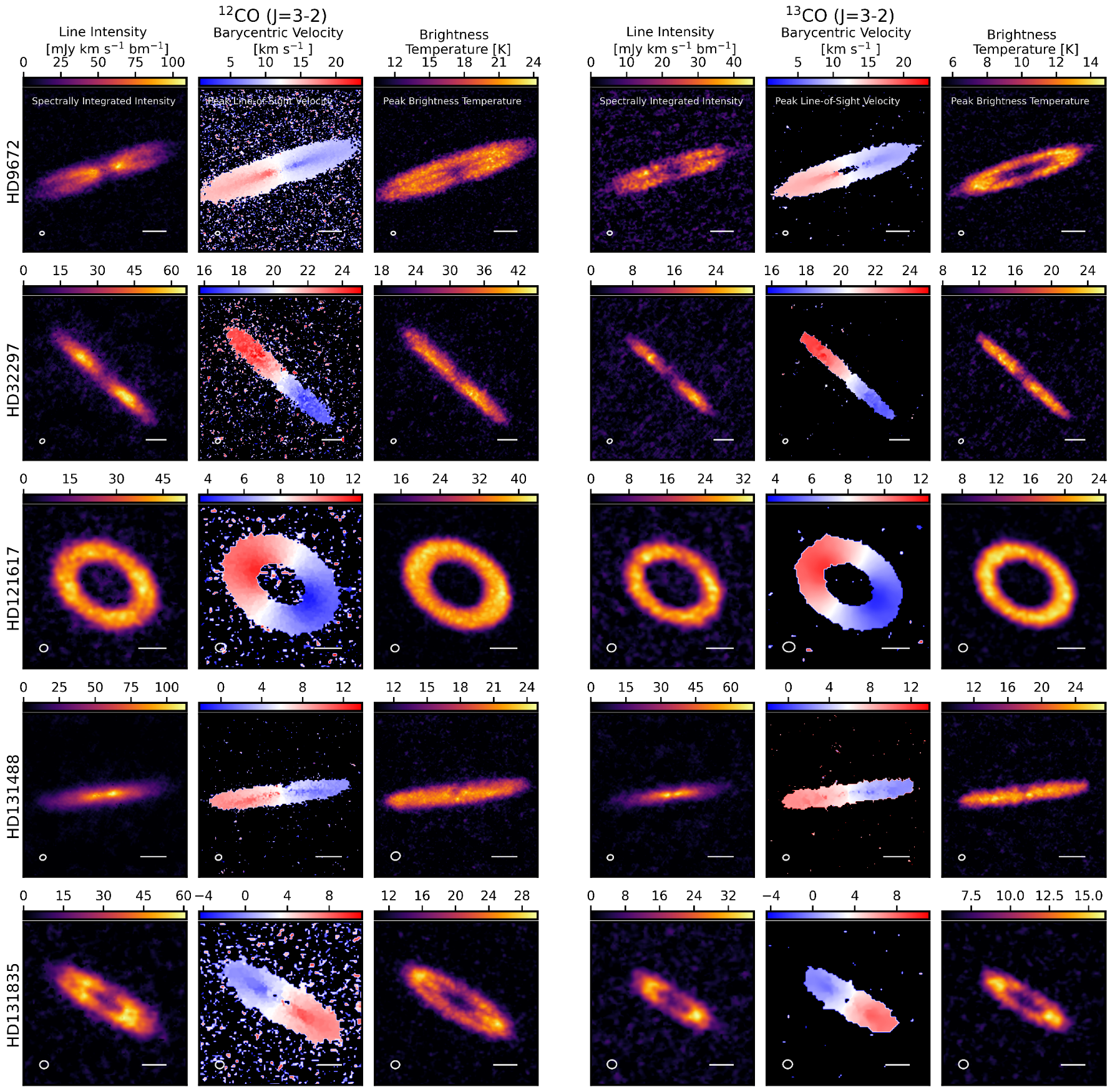}
    \caption{$^{12}$CO (left) and $^{13}$CO (right) emission of the five gas-bearing disks in ARKS,  HD~9672, HD~32297, HD~121617, HD~131488, and HD~131835 (from top to bottom). The first and fourth column in each row shows the spectrally integrated-intensity map of each disc for $^{12}$CO and $^{13}$CO, respectively. The minimum line intensity of these integrated-intensity maps was set at 0 mJy km s$^{-1}$ beam$^{-1}$. The second and fifth columns shows the peak velocity map of each disc for $^{12}$CO and $^{13}$CO, respectively. Each is centred on the stellar velocity of the systems, and any pixel outside of the velocity range shown in the colour bar is set to black. The peak-intensity (brightness temperature) map of each disc is shown in the third and sixth column for $^{12}$CO and $^{13}$CO, respectively. The minimum brightness temperature shown in the colour bar is the mean of the emission far from the discs. The ellipse in the lower left corners of the maps is the synthesized beam of the observations. The white scale bar in the bottom right corners indicates 50 au. }
    \label{12CO_gallery}
\end{figure*}

To best display the spectro-spatial structure of the ARKS gas-bearing sources, we created three different moment maps of each system; the line of sight peak velocity map (henceforth peak velocity map), the spectrally integrated-intensity map (henceforth integrated-intensity map) and the peak intensity/brightness temperature map. The peak velocity map displays the velocity field of the disc, in other words, the projected line-of-sight velocity (in the barycentric frame) of the gas in any given pixel. To calculate this, we fit a quadratic function to the peak and its two neighbouring channels in each of the pixels' spectra, and take this velocity of peak emission to describe the bulk velocity of the gas in each pixel. In the absence of a double-surface morphology typical of a protoplanetary disc ---such as that produced by CO freeze-out in the disc midplane---the velocity structure along each line of sight can typically be approximated by a single-peaked emission profile. In this case, a simple quadratic fit to the line profile at each pixel yields a reliable estimate of the bulk gas velocity. Using this quadratic fitting method has been shown to be better than using the intensity weighted average velocity method implemented by CASA for inferring line centroids \citep{bettermoments_teague}. 

The peak velocity maps for $^{12}$CO and $^{13}$CO are shown in the second and fifth column in Fig. \ref{12CO_gallery}, respectively. To ensure a good balance of reducing the noise from disc-free pixels while retaining velocity field information where the disc is faint, we clipped velocity values in pixels where the peak of the emission was detected at a significance below $3.5~\sigma$. We take the maximum of this quadratic to be the peak emission (in mJy beam$^{-1}$) in the spectrum of each given pixel. We converted the peak intensity maps to peak-brightness temperature, $T_{\rm B}$, units using Planck's law. This leads to the brightness temperature maps in Fig. \ref{12CO_gallery} (third and sixth columns), where a smoothing of 10 channels has been applied to the $^{12}$CO maps as a compromise between S/N and retaining as accurate a measure of peak $T_B$ as possible, though formally this leads to a slight underestimate of the brightness temperature.

Finally, we applied spectro-spatial shifting to the cube; this consists of shifting each pixel's spectrum so that its emission peak, originally located at the velocity shown in the peak velocity maps, was relocated to the stellar velocity. In particular, we spectrally integrated over all channels in the shifted cube where the disc was detected above $4~ \sigma$, which significantly reduced the number of channels containing disc signal to integrate over to create the integrated-intensity map of the discs. This consequently significantly increased the S/N of the final integrated-intensity maps. 

All three moment maps for each gas-bearing system are shown in Fig. \ref{12CO_gallery} for $^{12}$CO and $^{13}$CO, ordered by right ascension (RA) from top to bottom. We detect at high S/N and resolve $^{12}$CO and $^{13}$CO in each system over tens of resolution elements, with a variety of radial and vertical structure features seen in the integrated-intensity and the brightness temperature (peak intensity) maps, and the expected Keplerian velocity fields seen in the peak velocity maps.

\begin{equation}
    I \sim B_\nu = \frac{2h\nu^3}{c^2}\frac{1}{e^{\frac{h\nu}{kT}} - 1}
\end{equation}
\begin{equation}
    T = \frac{h\nu}{k}\frac{1}{\ln{(\frac{2h\nu^3}{Ic^2}+1)}}
\end{equation}

\subsection{Spectral and radial profiles}

\begin{figure*}[ht!]
    \centering
    \includegraphics[height=0.8\textheight, keepaspectratio]{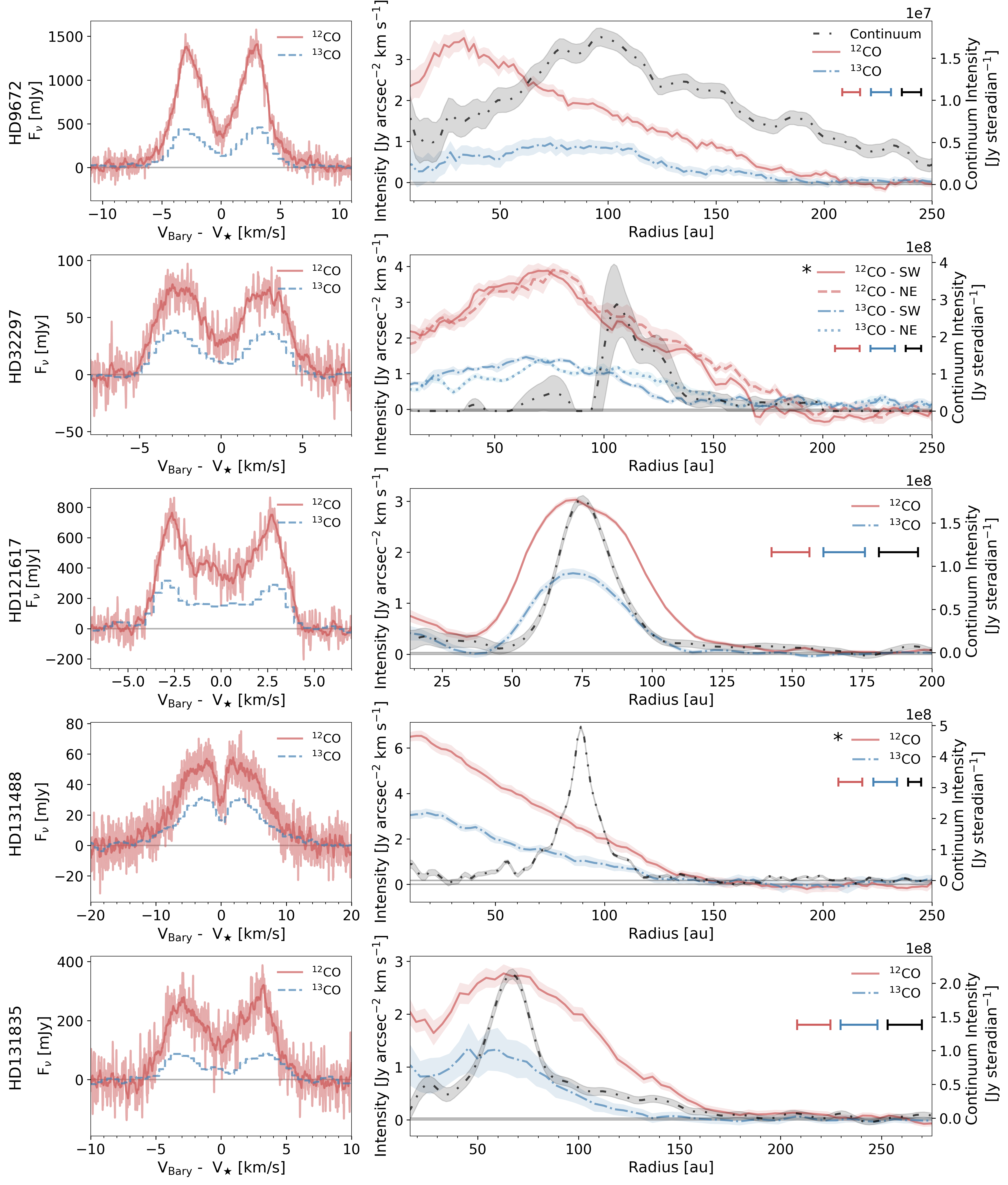}
    \caption{Left column: Spatially integrated spectrum of the $^{12}$CO(3-2) (red) and $^{13}$CO(3-2) (blue) emission from HD~9672, HD~32297, HD~121617, HD~131488, and HD~131835 in the barycentric velocity frame. The $^{12}$CO spectra are additionally shown smoothed with a running mean (window size = 20), chosen to remove noise variation while preserving spectral features. Right column: radial profiles for the same discs. Radial profiles for HD~9672, HD~121617 and HD~131488 were created by azimuthally integrating the emission in each disc. The radial profiles for HD~32297 and HD~131488 (marked by *) are surface brightness profiles as a function of projected separation from the star along the major axis (rather than as a function of radius). For HD~32297, the radial distribution varies for each side of the discs so both sides are shown, in both $^{12}$CO and $^{13}$CO. For HD~131488, there is no significant differences between both sides of the disc, so an average radial profile is shown. The black radial profiles show the CLEAN dust profiles \citep{Yinuo_arks}. In each panel, the minimum value of $R$ corresponds to the first resolution element of the observations.  The red, blue and black horizontal bars indicate the resolution of the observations. }
    \label{RadProf_gallery}
\end{figure*}

We created 1D spectra of $^{12}$CO and $^{13}$CO emission by spatially integrating the cube intensity between an inner and outer radius for each of the gas-bearing discs, using \verb|Gofish| \citep{rt_gofish}. We iteratively varied the radii until all detectable emission was included. The resulting spectra, which are shown on the left column of Fig. \ref{RadProf_gallery}, reveal the typical double peaked profile expected for a disc in Keplerian rotation around the central star. These spectra were shifted to the stellar velocity of each system - the stellar velocity was measured from the spectro-spatially shifted spectra, which will be discussed in Sect. \ref{observs_nondetect}.  
 
We also extracted azimuthally averaged intensity profiles of the $^{12}$CO and $^{13}$CO emission in each disc, seen on the right hand side of Fig. \ref{RadProf_gallery}. To create these profiles, we split the disc emission into concentric elliptical annuli, $\frac{1}{4}$ of the beam's major axis wide, each corresponding to a radial annulus in the orbital plane, accounting for the discs' inclination and position angle (Table \ref{tab:system_parameters}). For each radial annulus, we spectro-spatially stacked the emission by first shifting spectra in each of the annulus' pixels by the negative of their expected Keplerian line of sight velocity, and then averaging all the spectra from the annulus' pixels. We then integrated over the spectra to extract the intensity of the line at each radius, in mJy~beam$^{-1}$~km~s$^{-1}$, which we converted to mJy~arcsec$^{-2}$ km~s$^{-1}$ to allow direct comparison between discs. This was implemented through \verb|Gofish|'s  \verb|radial_profile| function. 

We find two of the gas-bearing discs, HD~32297 and HD~131488, to be too close to edge-on for the above technique to successfully produce radial profiles in the orbital plane; this is because for edge-on discs it is not possible to uniquely map an image pixel to an orbital radius (for a given inclination and PA). Therefore, we here present non-deprojected (on-sky) radial profiles (Fig. \ref{RadProf_gallery}). These were created by averaging spectrally integrated emission perpendicular to each disc's midplane, within a region $\pm$0\farcs3, 0\farcs2 perpendicular to the midplane of each disc, respectively, and taking the mean for each vertical stack of pixels. As there are significant differences in the emission between the two sides of HD~32297, which could indicate a potential asymmetry requiring confirmation, we plot the radial profiles from both sides of this edge-on disc for $^{12}$CO and $^{13}$CO in Fig. \ref{RadProf_gallery}. As there is no significant differences between both sides of HD~131488's disc, we plot the average radial profiles $^{12}$CO and $^{13}$CO in Fig. \ref{RadProf_gallery}. To measure the uncertainty on these radial profiles, the vertical averaging method was applied to many equally sized, $\pm$0\farcs3, 0\farcs2 slices of the image, in noise-only regions of the image to measure the noise in the background. This region-drawing and averaging (bootstrapping) was repeated many times in different parts of the image and the standard deviation of each of the average measurements was measured. Once the standard deviation of each of the average vertical measurements converged to a constant value for each vertical stack of pixels, it was taken as the error on each radial position in the radial profile. 

The resulting radial profiles of each disc are shown in the right column of Fig. \ref{RadProf_gallery}, and are compared to the continuum radial profiles from the same ARKS ALMA data \citep{Yinuo_arks}. Radial profiles comparing the gas distribution with the millimetre-sized dust grains observed with ALMA and with smaller, micron-sized dust grains observed in scattered light/NIR with SPHERE are presented in \cite{Julien_arks}.

\begin{figure}[ht!]
    \centering
    \includegraphics[height=0.79\textheight]{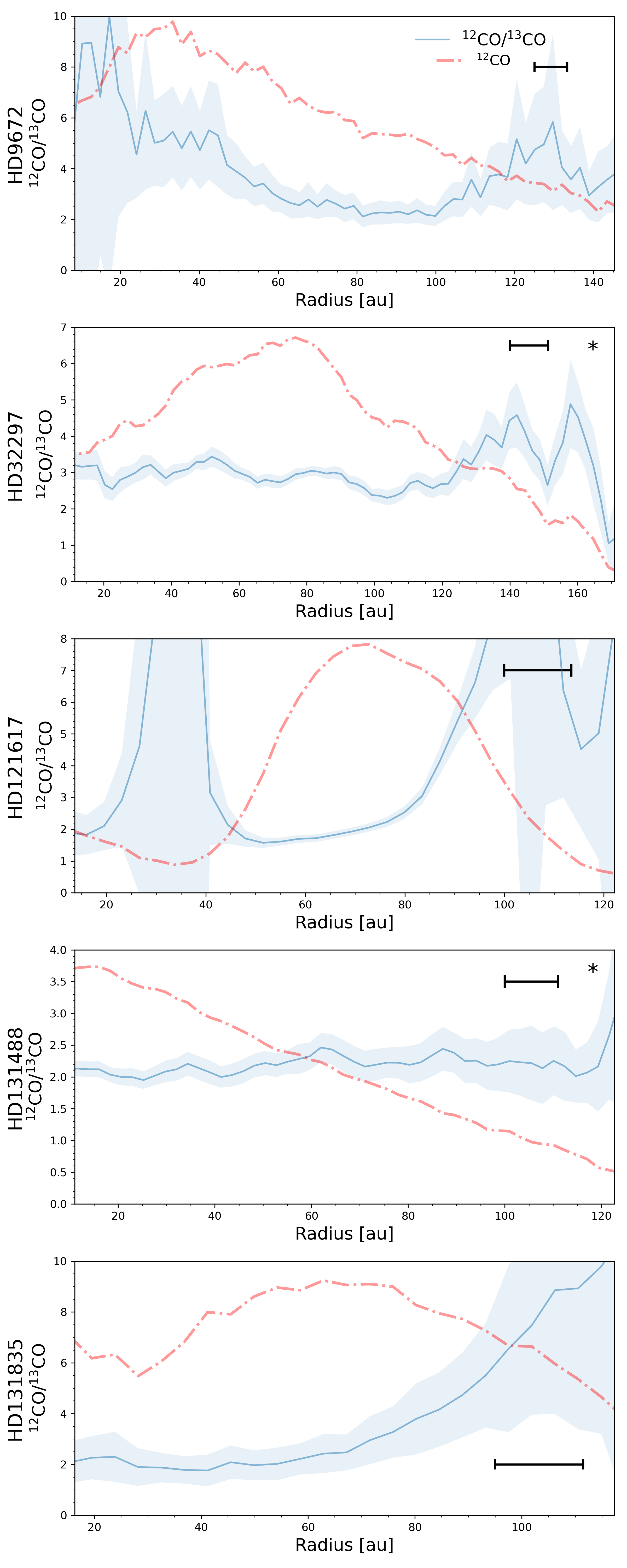}
    \caption{Radial profiles of $^{12}$CO/$^{13}$CO intensity ratios for HD~9672, HD~32297, HD~121617, HD~131488, and HD~131835 (blue). The $^{12}$CO radial profile is overplotted in red to show the gas distribution in each system. The $^{12}$CO/$^{13}$CO ratios for HD~32297 and HD~131488 (marked by an asterisk) were created from the average of the non-deprojected on-sky radial profiles. In each panel, the minimum value of R corresponds to the first resolution element of the observations. The black bar indicates the resolution of the observations.}
    \label{12_13_Ratio_gallery}
\end{figure}

As can be seen in Fig. \ref{RadProf_gallery}, in each system the $^{12}$CO gas appears to extend in all the way to the star. Thus, we can present an upper limit on the inner radius, $r_{\rm min}$, in each system, being half the resolution element (beam size) of our observations. These upper limits are in the few au range, from 4.2 au around HD~9672 to 8.3 au around HD~131488. We note that for the edge-on systems HD~32297 and HD~131488, this is based on the sky-projected profile and should not strictly be interpreted as an orbital radius.

To more accurately estimate an inner radius for all systems, we therefore present an alternative approach where we calculate the orbital radius corresponding to the highest velocity and on-sky along-midplane distance from the star at which a $3~\sigma$ detection is achieved anywhere in each system, accounting for the stellar mass and system inclination \citep[see Appendix C in ][]{matra2017exocometary_betapic}. This is applicable to systems with any inclination and leads to a likely better estimate for the inner $^{12}$CO radii of each system (though still formally an upper limit as fainter emission may extend further in) and is reported in Table \ref{tab:linefluxesofgas}.

Figure \ref{12_13_Ratio_gallery} shows how the $^{12}$CO/$^{13}$CO (spectrally integrated intensity) ratios change with radius for the gas-bearing sources. We only calculated this ratio in regions where both $^{12}$CO and $^{13}$CO are at least marginally detected; in other words, out to the radius at which point the $^{13}$CO radial profile is detected above the $2~\sigma$ level. This threshold was chosen as it gave the best compromise of showing the $^{12}$CO/$^{13}$CO ratio without displaying radii with high uncertainties due to the $^{13}$CO becoming undetected. Note that for the edge-on systems, the ratio radial profile is on-sky (non-deprojected) and was calculated for both sides of each disc. As we found no significant differences in the ratio radial profile between the two, in Fig. \ref{12_13_Ratio_gallery} we present an average of the two sides.

We find that the $^{12}$CO/$^{13}$CO intensity ratios at the radial $^{12}$CO intensity peaks of each system are $\sim 2.0 - 5.5$, and are well constrained with typical uncertainties of $\pm 6\% - 33\%$ at the peak of the $^{12}$CO radial profiles. Additionally, we do not find a strong dependence of the ratio with radius, within the fractional error, with ratios changing only by a factor $\sim 2-3$ across broad radial ranges of $\sim 120 - 160 $ au. The implications of these findings are discussed in Sect. \ref{discussion_isotopologue_ratios}.

\subsection{Spectro-spatial shifting and stacking}
\label{observs_nondetect}

Of the eighteen discs in the ARKS sample, thirteen were not previously known to host any gas. To investigate whether these discs have any detectable CO gas, we ran the first two \verb|tclean| rounds on each disc to create the dirty image cubes (with the robust parameter set to two to ensure the highest possible S/N) without any CLEAN deconvolution, as no signal was obviously present from visual inspection of the cubes.

To further boost the S/N and extract signals that would be too faint to see in each channel and in single resolution elements, we carried out spectro-spatial stacking \citep{Yen_2016, Teague_2016, matra2017detection_fomalhaut} using the \verb|GoFish| package \citep{rt_gofish}. To spectro-spatially stack the emission, we assigned a sky-projected line-of-sight velocity to every pixel under the assumption of Keplerian rotation and that the CO emission is co-located with the dust emission. Given the known disc inclination, position angle (PA), both derived from MCMC sampling carried out by \citet{SebasARKSpaper}, and stellar mass, estimated as described in Appendix A in \citet{SebasARKSpaper}, the spectrum in each pixel was then shifted by the negative of its assigned sky-projected (line of sight) Keplerian velocity to the spectral (stellar) velocity (using literature values as listed in Table \ref{tab:beamsizes_gasrich}). The minimum and maximum orbital radii of pixels to be shifted and stacked were selected by $r_{min/max}$ as in Eq. \ref{rminmax}. For the edge-on systems, as mentioned in Sect. \ref{observation_sec}, it is not possible to assign a Keplerian velocity uniquely to each pixel. Therefore, we conservatively set $r_{min}=0$ to ensure that the radial selection includes all on-sky pixels that are likely to contain emission. For the gas-free systems, the spectro-spatial shift was repeated twice, once assuming an orbital inclination $i$ (from the known geometry of dust emission) and once with inclination $i+180^\circ$, to account for the two possible rotation directions of the disc.

\begin{figure}[h]
    \centering
    \includegraphics[width=0.99\linewidth]{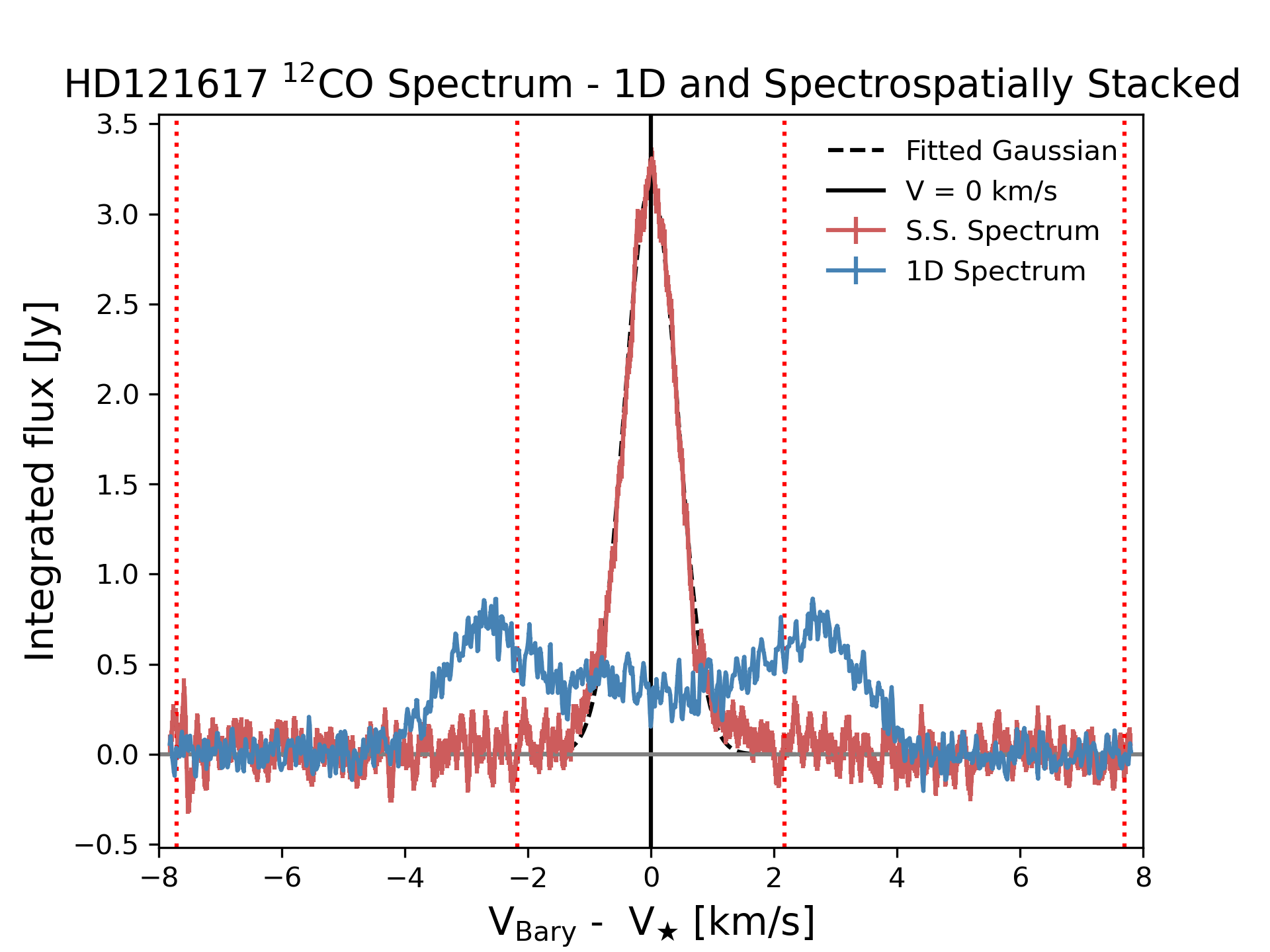}
    \caption{ $^{12}$CO spatially integrated spectrum of HD~121617(blue) and its spectro-spatially stacked (S.S.) spectrum (red). The dotted red lines indicate the velocities in which integration was carried out to calculate the error on the line flux measurement. $V_*$ represents the stellar velocity of HD121617, determined by fitting a Gaussian profile to the spectro-spatially shifted spectrum.}
    \label{HD~121617_SSS_1Dspec}
\end{figure}

Once all the spectra in the different pixels are aligned in velocity, they can be summed to significantly boost the S/N, as demonstrated for the gas-bearing disc HD~121617 in Fig. \ref{HD~121617_SSS_1Dspec}. This shows the spectrum integrated spatially over pixels that correspond to orbital (as opposed to on-sky) radii between $r_{min}$ and $r_{max}$, as extracted directly from the image cube of HD~121617 in blue. We show the same spectrum following the spectro-spatial stacking (S.S. spectrum) process in red. Thus, if gas is present in the previously undetected systems, the stacked profile will present a single peak at the stellar velocity rather than a Keplerian profile, with the same noise level, leading to a significant peak S/N boost, as demonstrated in Fig. \ref{HD~121617_SSS_1Dspec}.

The spectro-spatially shifted and stacked spectra for each of the 13 systems where no CO was previously detected are shown in Fig. \ref{gasfree_gallery} (two spectra plotted to represent each rotation direction). Despite application of the S/N-boosting spectro-spatial stacking technique, we do not detect $^{12}$CO in any of the systems.

For the gas-bearing disks, to measure the barycentric velocity of each system, we fitted a Gaussian to the spectro-spatially stacked spectra such as shown for HD121617 in Fig. \ref{HD~121617_SSS_1Dspec}. Although Gaussians are not a perfect fit to the detailed shape of the single peaks, they are sufficient to determine the velocity centroid of the emission.

These velocities are presented in Table \ref{tab:system_parameters} and are consistent with the stellar velocities used in the mask making process. Note that the uncertainties listed on these velocities are just statistically based from the Gaussian fit and do not include possible systematic errors causes by imperfect shifting or asymmetric disc velocity fields. As a check of the validity of the velocities measured with this method, we compared to the stellar velocity  of HD~121617 to the stellar velocity measured by \cite{Aoife_arks} for the same star ($V_* = 7858^{+2}_{-3} $m~s$^{-1}$) through fitting a Keplerian model to the peak velocity map. We found the two measured velocities to be consistent within $3~\sigma$ of each other. Additionally, we checked the stellar velocity of HD131488 measured through high-resolution near-IR CRIRES+ spectroscopy \citep[$V_* = 5.860^{+0.096}_{-0.012}$ ~km~s$^{-1}$,][]{Kevin_H2_CO}, and found it to be also consistent with our measurement.

\subsection{Flux and mass lower limit measurements}
\label{res:flux_mass}

To measure the line flux of the $^{12}$CO and $^{13}$CO lines in each of the three non-edge-on gas-bearing discs, we used the spectro-spatially shifted and integrated spectra described in Sect. \ref{observs_nondetect}. For example, the red spectrum in Fig. \ref{HD~121617_SSS_1Dspec} was used to measure the integrated line flux of HD~121617. This way, we can take advantage of the narrower line width (for the same spectral RMS before and after spectro-spatial stacking) to obtain a lower uncertainty on the velocity integrated line flux. 

We integrated over the red, spectro-spatially stacked spectrum created and shown in Fig. \ref{HD~121617_SSS_1Dspec}. To include all the emission while avoiding unnecessary noisy regions, we used the standard deviation sigma from the Gaussian fit described in Sect. \ref{observs_nondetect} to guide the limits for the integration. We set the limits of integration to $V_{*} \pm 5~\sigma$. We then visually confirmed that this covered the entirety of the line emission from each disc. 

We use bootstrapping to calculate the error on these line flux measurements. To do so, we repeat the integration to calculate the error on this line flux measurement a large number (>100) of times over the exact same number of channels as the actual measurement, but in noise-only regions of the spectro-spatially stacked spectrum. We choose the standard deviation of these bootstrap integrations as the statistical uncertainty on our line flux measurement. The number of iterations over which the integration was repeated was set by visually inspecting the standard deviations of the integration and was accepted as the error once it had converged to a constant value, making sure to use a sufficient number of noisy regions to allow this standard deviation value to converge. This statistical uncertainty was then added in quadrature to the systematic flux density calibration error associated with ALMA band 7 observations, which is 10\% of the measured flux \citep{ALMAHandbook}; we find that the flux calibration systematic dominates the uncertainty for all discs. 

For the edge-on gas-bearing systems, this method of measuring the line flux was ineffective, as due to their edge-on inclination, emission above and below the midplane was missed in the shifting and stacking process. Thus, to measure the integrated line flux for the two edge-on systems, we simply spatially integrated the intensity over the region where disc emission is present. For HD~32297 and HD~131488 respectively, the regions chosen were (78 $\times$ 500) au and (61 $\times$ 500) au boxes with the long sides along the midplane and short sides perpendicular to it, centered on their respective stellar positions. We obtained the integrated flux uncertainty by multiplying the RMS in the spectrally integrated intensity maps (in units of Jy km s$^{-1}$ beam$^{-1}$) by the square root of the number of independent resolution elements (beams) in the respective boxes, and added this in quadrature to the systematic absolute flux calibration error, where the latter dominates the error budget. The integrated line fluxes of $^{12}$CO and $^{13}$CO for gas-bearing discs are presented in Table \ref{tab:linefluxesofgas}.

We used Eq. 2 in \cite{matra2015co_fomalhaut} to calculate the $^{12}$CO optically thin ($\tau_{^{12}\rm CO} <<1 $) gas mass in each disc using the integrated line fluxes. The calculation requires knowledge of the upper level (in our case, J=3) population of the CO molecule. For simplicity, we calculate this assuming local thermodynamic equilibrium (LTE) and a CO excitation temperature equal to the peak-brightness temperature measured from the $^{12}$CO brightness temperature maps. If the gas is optically thin, this will 
lead to an underestimation of the excitation temperature (as $T_{\rm B} < T_{\rm exc}$), but that would mean that the masses presented in Table \ref{tab:linefluxesofgas} remain lower limits as higher $T_{\rm exc}$ would lead to higher derived masses. These values (Table \ref{tab:linefluxesofgas}) are of order $\sim1-6\times10^{-4}$ M$_{\oplus}$. The relationship between this optically thin mass and the one that would be derived for any optical depth value is 

\begin{gather}
\label{eqn:massthick}
    M_{{^{12}\rm CO, thick}} = \frac{\tau_{^{12}\rm CO}}{1-e^{-\tau_{^{12}\rm CO}}} M_{{^{12}\rm CO, thin}} \hspace{1cm} ,
\end{gather}

with the caveat that this is in the simple assumption of a disc of spatially uniform CO density and temperature. This allows us to also report a mass in the assumption that the $^{12}$CO is optically thick with an assumed optical depth of 77 (Table \ref{tab:linefluxesofgas}), motivated our findings in Sects. \ref{discussion_isotopologue_ratios} - \ref{dis:brightness_temp}, showing that the $^{12}$CO/$^{13}$CO ratio radial profiles are likely explained by both isotopologues being optically thick everywhere they are detected in the majority of systems. Then, if $^{13}$CO is optically thick, its optical depth is at least 1. If the ISM $^{12}$C/$^{13}$C abundance ratio of 77 applies, that implies that the $^{12}$CO optical depth must be at least 77 \citep{wilson1994abundances}. As such, our mass derived in the assumption of a $^{12}$CO optical depth of 77 is to be considered a lower limit in a scenario where both $^{12}$CO and $^{13}$CO are optically thick, which we deem likely for most detected systems.

\begin{table*}[]
\centering
\caption{Upper limits on the inner radii of $^{12}$CO, integrated line flux, and corresponding gas masses of $^{12}$CO(J=3-2).}
\label{tab:linefluxesofgas}
\begin{tabular}{cccccc}
\hline
 & \begin{tabular}[c]{@{}c@{}}$r_{\rm min ^{12}CO}$ \\ {[}au{]}\end{tabular}  &\begin{tabular}[c]{@{}c@{}}$^{12}$CO (J=3-2)\\ Integrated Line Flux\\ {[}Jy~km~s$^{-1}${]}\end{tabular} & \begin{tabular}[c]{@{}c@{}}$^{13}$CO (J=3-2)\\ Integrated Line Flux\\ {[}Jy~km~s$^{-1}${]}\end{tabular} & \begin{tabular}[c]{@{}c@{}}$^{12}$CO (J=3-2)\\ $\tau_{^{12}\rm CO} <<1$ Mass \\{[$10^{-4} M_\oplus$]}\end{tabular} & \begin{tabular}[c]{@{}c@{}}$^{12}$CO (J=3-2)\\ $\tau_{^{12}\rm CO} =77$ Mass \\{[$10^{-2} M_\oplus$]}\end{tabular} \\ \hline \hline 
HD~9672 & $ 3$ & 8.9 $\pm$ 0.9 & 3.0 $\pm$ 0.3 & $2.8 \pm 0.3 $ & $ 2.2 \pm 0.2$ \\
HD~32297 & $68$ & 3.7 $\pm$ 0.4 & 1.3 $\pm$ 0.1 & $5.8 \pm 0.6 $ & $4.5\pm 0.5$ \\ 
HD~121617 & $28$ & 3.6 $\pm$ 0.4 & 1.5 $\pm$ 0.2 &  $4.6 \pm 0.5 $& $3.5 \pm 0.3 $ \\ 
HD~131488 & $6$ & 2.7 $\pm$ 0.3 & 1.3 $\pm$ 0.1 & $5.9 \pm 0.7$ & $4.5 \pm 0.5 $ \\ 
HD~131835 & $21$ & 1.9 $\pm$ 0.2 & 0.61 $\pm$ 0.06 & $ 2.9 \pm 0.3  $ & $2.2 \pm 0.2 $ \\ \hline
\end{tabular}
\caption*{\textbf{Notes:} Upper limits on $r_{\rm min}$ correspond to the highest velocity and on-sky along-midplane distance from the star at which a $3~\sigma$ detection is achieved anywhere in each system. Integrated line fluxes were measured by spectro-spatially stacking the spectra of the ARKS gas-bearing targets. $^{12}$CO gas masses derived assuming $^{12}$CO are optically thin ($\tau_{{^12}\rm CO} << 1)$ and assuming that the $^{12}$CO and $^{13}$CO is optically thick, and the $^{12}$CO/$^{13}$CO ratio is 77 (ISM-like). The optically thin gas mass was calculated using Eq. 2 in \cite{matra2015co_fomalhaut} and by taking the peak-brightness temperatures from the peak-brightness temperature maps. Optically thin masses were scaled by the ISM $^{12}$CO/$^{13}$CO ratio to calculate the optically thick mass. }
\end{table*}

For the discs with non-detected gas, we calculated upper limits of the $^{12}$CO integrated line flux by using the same bootstrapping method as described above, applied to the spectro-spatially shifted spectra in Fig. \ref{gasfree_gallery}. We spectrally integrated over 7 km~s$^{-1}$ wide noisy regions, located in random regions of each spectrum, and measured the standard deviation of the bootstrapped samples to derive the uncertainty $\sigma$ on the spectrally integrated flux of any line, if present. We chose 7 km~s$^{-1}$ because the gas-bearing spectro-spatially stacked spectra were on average 7 km~s$^{-1}$ wide across a range of disc radial profiles and inclinations. We used this to set 3 $\sigma$ upper limits on the line flux for all our undetected discs, presented in Table \ref{tab:gas_free_discs_line_flux}. Given that $^{12}$CO is not detected, we do not report upper limits for the rarer $^{13}$CO isotopologue, as it would not be detected either.

\section{Discussion}
\label{discussion_sec}

\subsection{Radial CO distribution}
\label{discussion_radial_co_dist}
With the high spatial resolution ARKS delivers, we were able to extract resolved radial profiles for the gas-bearing discs. Remarkably, we find that in all systems the CO emission is significantly broader and extends much closer to the star compared to the dust emission. Of the five gas-bearing discs in ARKS, only HD~121617 has a ring-like appearance, though even this disc has gas interior to the dust ring, with emission rising again inwards of a radial minimum at $\sim 34$ au (Fig. \ref{RadProf_gallery}). Interestingly, in each of the gas-bearing discs, the CO intensity distribution peaks interior to the millimetre-sized dust grains observed within ARKS. This could be an optical depth effect if the gas in each disc is optically thick, or an indication that the gas release is modulated by the temperature of solids which is higher closer to the star \citep{bonsor2023secondary}. In three discs, HD~9672, HD~32297, and HD~131488, we see large differences ($> 40$ au) in the peak locations between the CO and the dust (Fig. \ref{RadProf_gallery}). In HD~121617 and HD~131835, we see smaller, but still significant, offsets between the gas and the millimetre-sized dust grains.

This raises the question of whether CO can be continuously produced in a second-generation scenario, given it is not entirely co-located with the dust \citep{kospal2013alma}. However, gas released from exocometary ice is expected to viscously spread radially inwards and outwards of the belt \citep{kral2016self}. For $^{12}$CO to have time to spread (as opposed to the long-lived atomic dissociation products), significant amounts of shielding would need to be in place. This is because for CO gas to persist past $\sim 130$ years (the unshielded CO photodissociation timescale due to interstellar FUV photons) a shielding agent such as H$_2$, C, or CO itself is required \citep{heays2017photodissociation, matra2015co_fomalhaut}. The fact that we observe the CO to be broad/spreading throughout these systems means it must be shielded in the vertical direction from the Interstellar Radiation Field (ISRF) and in the radial direction from stellar UV photons, as in regions sufficiently close to the star, stellar photons dominate the UV field and photodissociation processes. In order for the gas to persist and inwardly propagate in this region, its shielding against the stellar UV photons must also be efficient enough (especially for HD~9672 and HD~131488, where the CO gas is present very close to the luminous star). This shielding could be done by CI and/or CO self-shielding \citep{marino2020population}. Therefore, in a secondary-generation scenario, the observed broad CO radial profiles of these CO-bearing discs can only be explained if the CO and/or CI column densities are sufficiently high to provide significant shielding or the viscosity of the gas is sufficiently high to allow the gas to spread within its photodissociation timescales. On the other hand, the CO could be primordial in origin, with large amounts of unseen H$_2$ shielding it. Whether the gas is primordial or secondary in HD~121617 is further discussed in \citet{Aoife_arks}, \cite{hd121617_arks}, and \cite{Phillip_arks}. 

\subsection{Resolved isotopologue ratios and optical depth}
\label{discussion_isotopologue_ratios}

The $^{12}$CO/$^{13}$CO intensity ratio was radially resolved in each of the gas-bearing discs and are found to be between $\sim2.0 - 5.5$ at the peak $^{12}$CO locations, which also corresponds to the peak location for $^{13}$CO in all systems but HD~9672 (Figs. \ref{RadProf_gallery} and \ref{12_13_Ratio_gallery}). These low peak or otherwise disc-averaged $^{12}$CO/$^{13}$CO line ratios were previously interpreted to indicate optically thick $^{12}$CO as they are much lower than the expected values from the ISM, ${^{12}\text{CO}}/ {^{13}\text{CO}} = 77$ \citep{wilson1994abundances}. Interestingly, we find most ratio profiles in Fig. \ref{12_13_Ratio_gallery} to be consistent with being radially flat across the entire radial region where $^{13}$CO is detected at the $2\sigma$ level (note that the shaded region shown in Fig. \ref{12_13_Ratio_gallery} is $\pm1\sigma$). This radial change of the isotopologue ratio can be interpreted in terms of the optical depth of the five gas-bearing systems.

If we assume that optically thick $^{12}$CO is co-located with optically thin $^{13}$CO, that the CO is in local thermodynamic equilibrium (LTE), and that $T = T_{^{12}\rm CO} = T_{^{13}\rm CO}$, we could write the specific intensity of the line describing the spectrum at a specific radius as 

\begin{gather} 
 I_{\nu_{^{12}\rm CO}} \simeq S_\nu (T) , \\
 I_{\nu_{^{13}\rm CO}} = S_\nu (T)(1 - e^{- \tau_{^{13}\rm CO}})   ,
\end{gather}

where the source function is equal the Planck function, $S_\nu(T) = B_\nu (T)$, as the gas is in LTE and $\tau_{^{13}\rm CO}$ is the optical depth of the $^{13}$CO. This assumption is based on previous studies where the $^{12}$CO was calculated to be very optically thick in all discs but where $^{13}$CO was assumed to be optically thin \citep[e.g. HD~9672, HD~121617;][]{moor2017, moor2019new}. This relation still holds if we do not assume LTE for the co-located $^{12}$CO and $^{13}$CO if they are subject to the same excitation conditions. Accordingly, we arrive at the following expression for the specific intensity ratio: 
 
\begin{gather}
\label{eqn_ratio}
 \frac{I_{\nu_{^{12}\rm CO}}}{I_{\nu_{^{13}\rm CO}}} = \frac{1}{1 - e^{- \tau_{^{13}\rm CO}}}  \sim \frac{1}{\tau_{^{13}\rm CO}} 
\end{gather}

\noindent as $\tau_{^{13}\rm CO} << 1$ for optically thin $^{13}{\rm CO}$. We know that the optical depth is proportional to the column density of the gas, $\tau_{^{13}\rm CO} \propto N_{^{13}\rm CO}$, and therefore that the $^{13}$CO specific intensity is also proportional to the column density, $I_{\nu_{^{13}\rm CO}}\propto N_{^{13}\rm CO}$. Thus, if we see a decrease radially in optically thin $^{13}$CO specific intensity from a peak, this indicates a decrease in $^{13}$CO optical depth and column density; in turn, from Eq. \ref{eqn_ratio} above we should see an increase in the $^{12}$CO/$^{13}$CO specific intensity ratio. Overall, for optically thick $^{12}$CO and optically thin $^{13}$CO, a decrease in $^{13}$CO intensity should be accompanied by an increase in the $^{12}$CO/$^{13}$CO specific intensity line ratio. 

We highlight that in Fig. \ref{RadProf_gallery} we are showing the spectral integral of specific intensity (i.e. the intensity $I$) rather than the specific intensity $I_{\nu}$ discussed above, and that the line ratio profiles in Fig. \ref{12_13_Ratio_gallery} also show ratios of intensities rather than specific intensities. However, a similar behaviour as described above would be expected for intensity ratios as for specific intensity ratios, as the integral of an optically thick line changes much more slowly than an optically thin line. In conclusion, this indicates that the flat intensity ratios with radius shown in Fig. \ref{12_13_Ratio_gallery} rule out that the $^{12}$CO is optically thick and co-located with optically thin $^{13}$CO.  

Radially flat $^{12}$CO/$^{13}$CO ratios can thus be explained if both isotopologues are optically thick everywhere they are significantly detected, as inferred by \citet{Aoife_arks} for HD~121617. In this case, both $^{12}$CO and $^{13}$CO specific intensities (and also integrated intensities) would trace the radial temperature distribution, which is likely to be monotonically decreasing with radius (e.g. $^{12}$CO for HD~9672, and both isotopologues for HD~131488). It is important to note that if both isotopologues were optically thick, the underlying abundance ratios would be completely unconstrained. On the other hand, it remains formally possible that for discs other than HD~121617 both isotopologue lines could be optically thin everywhere they are detected. However, it is not clear what physical mechanism could be responsible for such a large change in the abundance ratio from the ISM abundance ratio, if both isotopologues are optically thin. 

Thus, to explain our largely radially constant (within $\pm 1 \sigma$) $^{12}$CO/$^{13}$CO ratios, both isotopologues have to be either optically thick, or optically thin, at every location where they are detected. To break this degeneracy, radiative transfer modelling needs to be carried out, as in \citet{Aoife_arks} for HD~121617. 

Interestingly, \cite{zhang2021molecules} found the $^{12}$CO/$^{13}$CO ratio to be $\sim 2 - 4$ (see their Fig. 2, top two rows) in the gas-rich protoplanetary discs observed in the ALMA MAPS large program, which they interpreted to be due to the optical thickness of the CO gas. Combining this with the results of \citet{Aoife_arks}, it is likely that we are observing the same phenomenon with our comparably low  $^{12}$CO/$^{13}$CO ratios being caused by optically thick gas in each of the gas-bearing discs included in this study. 

\subsection{Temperature and CO mass estimates}
\label{dis:brightness_temp}

If $^{12}$CO and $^{13}$CO are optically thick, the peak intensity and therefore brightness temperature is a direct probe of the excitation temperature, which if in LTE, also corresponds to the gas kinetic temperature. Interestingly, we measure peak-brightness temperatures of $\sim 25 - 50 $ K from the $^{12}$CO peak-intensity/brightness temperature maps (Fig. \ref{12CO_gallery}). Each measured peak-brightness temperature is lower than the corresponding blackbody temperature at the discs' peak radial location (which range from $54 $ K $< T_{\rm BB} < 137$ K), which is consistent with what was found with lower resolution observations of gas-bearing debris discs \citep[e.g.][]{moor2019new}. 

This difference in temperature could be caused by several mechanisms: the first possibility is that LTE does not apply and thus the kinetic temperature might be higher and closer to the black body temperature. This could be caused by low gas densities (e.g. in the absence of $H_2$) in which case, these temperatures could be lower limits to the gas kinetic temperature. The second possibility is that LTE does apply but differing heating/cooling processes (decoupling) of gas from dust is causing the difference. 

With co-located optically thick $^{12}$CO and $^{13}$CO, we could naively expect the measured peak-brightness temperatures of both isotopologues would be equal (as we would expect $T_{^{12} \rm CO} = T_{^{13} \rm CO} = T_{\rm exc} = T_{\rm B}$). However, as can be seen when comparing between the peak-brightness temperatures for $^{12}$CO and $^{13}$CO for the majority of the discs (Fig. \ref{12CO_gallery}), there is a factor of $\sim 2$ difference between the peak-brightness temperatures between $^{12}$CO and $^{13}$CO. This is likely due to the much lower spectral resolution in the $^{13}$CO observations; the saturated line was convolved by the lower resolution element, and thus the resulting observed peak intensity/brightness temperature would be lower that the true peak intensity/brightness temperature. An example that potentially showcases this issue is HD~131488, in which the local line profiles are wider and therefore better resolved, likely because the gas extends closer to the star where broadening by Keplerian shear is more significant \citep[][Appendix A]{Aoife_arks}; the $^{13}$CO line was resolved and in this disc, we observe comparable peak temperatures for both $^{12}$CO and $^{13}$CO.  

As many of the $^{12}$CO/$^{13}$CO intensity radial profiles can be explained by both isotopologues being both optically thick or both optically thin, in Sect. \ref{res:flux_mass} (Table \ref{tab:linefluxesofgas}) we reported $^{12}$CO masses for both scenarios. In the optically thick case however, it is not formally possible to directly determine CO gas masses from line fluxes, so we calculated an optically thick mass assuming $\tau_{^{12}\rm CO}=77$. This assumption is based on the fact that if $^{12}$CO is optically thick, then $^{13}$CO must be as well ($\tau_{^{13} \rm CO}>>1$), as demonstrated in Sect. \ref{discussion_isotopologue_ratios}. Assuming ISM abundance ratios \citep[$^{12}$CO/$^{13}$CO = 77,][]{wilson1994abundances}, $\tau_{^{12}\rm CO} = 77\tau_{^{13}\rm CO}$, and therefore $\tau_{^{12}\rm CO} >> 77$. Therefore, our reported $\tau_{^{12}\rm CO}=77$ masses, while giving a more realistic idea of the $^{12}$CO optical depth if both isotopologues are thick, should be interpreted as lower limits, as $\tau_{^{13}\rm CO}$ is potentially much higher than 1.

\subsection{$^{12}$CO (J=3-2) population study}
To compare the ARKS sample to the broader population, we combine ARKS line fluxes (including 13 upper limits) with any archival discs observed with ALMA with a reported line flux for the $^{12}$CO J=3-2 line. We also include archival $^{12}$CO J=2-1 ALMA observations by scaling the $^{12}$CO J=2-1 line flux by the average of the $^{12}$CO J=3-2/J=2-1 ratio ($=1.67$) of all discs with both lines observed.

\begin{figure}[h]
    \centering
    \vspace{-2mm}
    \hspace{-9mm}
    \includegraphics[width=1.1\linewidth]{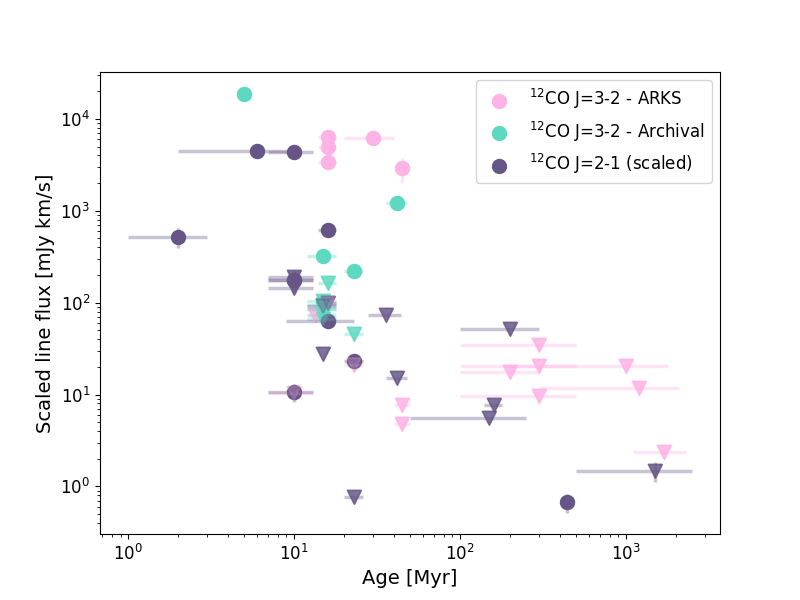}
    \vspace{-8mm}
    \caption{Line flux of $^{12}$CO (J=3-2) scaled to 100 pc as a function of the stellar age. The pink points show $^{12}$CO (J=3-2) fluxes measured in this work, the green points show archival $^{12}$CO (J=3-2) fluxes, and the purple points show archival $^{12}$CO (J=2-1) fluxes that were scaled by the average $^{12}$CO J=3-2/J=2-1 ratio of all discs for which both lines were observed (1.67). The archival sources and the corresponding references are listed in Table \ref{tab:archival_linefluxes}. The circles represent resolved data points, and the triangles represent upper limits on $^{12}$CO.}
    \label{scaledfluxage}
\end{figure}

We scaled all line flux measurements to 100 pc to compare the intrinsic line luminosity of all discs. Figure \ref{scaledfluxage} shows the scaled line fluxes as a function of system age. We see that gas detections are mostly found in systems younger than 50 Myr, though Fomalhaut \citep[$440 \pm 40$ Myr, ][]{mamajek2012age} and $\eta$ Corvi (HD~109085, $1500^{+1000}_{-500}$ Myr, \cite{pearce2022planet} and references therein) are notable exceptions. Although the figure shows there is a clear trend of detection rate with age, no trend is present within detected systems, which could be due to a large scatter caused by different system properties and/or evolution pathways. 

\begin{figure}[ht!]
    \centering
    \vspace{-5mm}
    \hspace{-9mm}
    \includegraphics[width=1.1\linewidth]{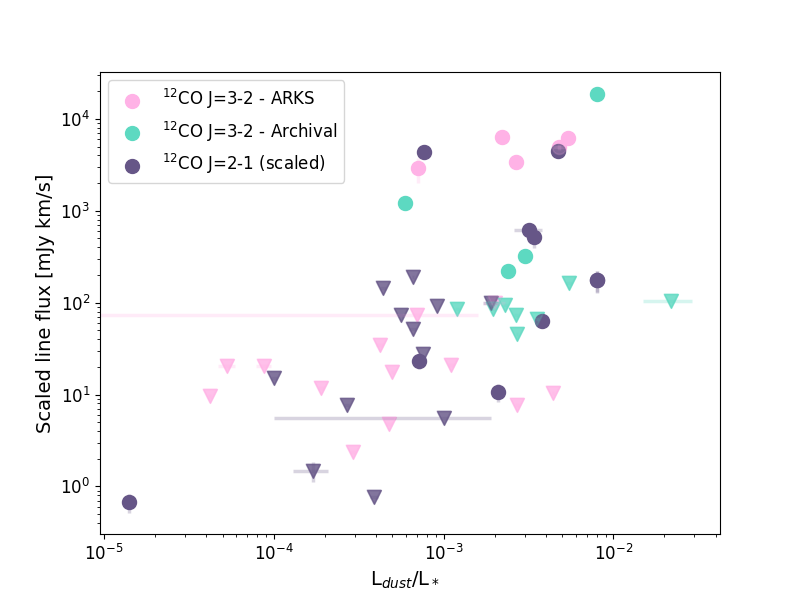}
    \vspace{-8mm}
    \caption{Similar to Fig. \ref{scaledfluxage}, but for the line flux of $^{12}$CO scaled to 100 pc as a function of the fractional luminosities (L$_{dust}$/L$_*$) of the systems.} 
    \label{scaledfluxfraclum}
\end{figure}

In Fig. \ref{scaledfluxfraclum}, we show the relationship between the scaled line fluxes and the fractional luminosity of the discs. We observe a positive trend in which discs with higher fractional luminosities correspond to the gas-bearing debris discs implying that systems in which the most dust is being released/created corresponds to the systems with gas in the discs. This could indicate a secondary origin for the gas in these systems as the CO gas could originate in the same collisional cascade that creates the dust \citep{matra2019ubiquity_TWA7}. However, this could simply be a consequence of the age trend, given that younger systems will typically also have larger fractional luminosities due to collisional evolution \citep[e.g.][]{wyatt2007steady}. Nonetheless, the presence of upper limits for the highest fractional luminosities (and youngest ages) implies there is a large amount of scatter for dusty/young systems, which again could be caused by different system properties and/or evolutionary pathways \citep{marino2020population}. 

\subsection{Individual discs} 

For a discussion on each individual disc, their unique CO distributions and comparisons with previous, low resolution observations, please see Appendix \ref{App:Indiv_discs}.

\section{Conclusions}
\label{conclusions_sec}

We presented the ALMA $^{12}$CO and $^{13}$CO observations of 18 debris discs in the ARKS sample. Our conclusions are listed below.

\begin{itemize}
    \item The distributions of $^{12}$CO and $^{13}$CO are observed to be broader than the continuum in each of the five gas-bearing discs in our sample. In each disc, the CO peaks interior to the dust, although the offset between the dust and gas varies with each system. The radial distribution for CO is much broader, and we detected CO that extended in as far as the inner resolution element of the observations in all systems. We calculated velocity-based upper limits to the inner disc radii of $3 - 68$ au.
    \item We presented radially resolved $^{12}$CO/$^{13}$CO isotopologue flux ratios for the five gas-bearing discs in our sample. This allowed us to infer the optical depth of the CO. Based on the $^{12}$CO/$^{13}$CO ratios that are consistent with being radially flat at the $3 \sigma$ level presented for HD~32297, HD~121617, HD~131488, and HD~131835, we rule out that $^{12}$CO is optically thick and co-located with optically thin $^{13}$CO. We instead conclude that the isotopologues are either both optically thick or both optically thin in the discs. HD~9672 alone shows different radial profiles for $^{12}$CO and $^{13}$CO, which could imply optically thick $^{12}$CO with optically thin $^{13}$CO. This is further supported by the shape of its global spectral line profile. 
    \item We measured the integrated line flux of $^{12}$CO and $^{13}$CO in each gas-bearing system. Based on this, we calculated the optically thin and optically thick masses of $^{12}$CO. The optically thick masses should be considered a strict lower limit because they were calculated by assuming ISM ratios of $^{12}$CO/$^{13}$CO and that $\tau_{^{12}\rm CO} = 77$, which is likely an underestimate of $\tau_{^{12}\rm CO}$, and a higher $\tau_{^{12}\rm CO}$ would increase the optically thick mass we calculated. 
    \item We conducted a deep search for $^{12}$CO in the 13 gas-free discs in the ARKS sample and did not detect CO in any of them. This allowed us to set strict upper limits on the integrated line flux in these 13 systems. 
    \item We combined ARKS-derived CO J=3-2 line luminosities (or upper limits) with literature values to confirm that gas is preferentially detected in younger systems and/or in systems with a high fractional luminosity, but we highlight the large scatter in the line luminosities within the cohort of systems that are younger or have a higher fractional luminosity. This could potentially be caused by a range of host star properties and/or evolutionary pathways. 
\end{itemize}

\begin{acknowledgements}

SMM and  LM acknowledge funding by the European Union through the E-BEANS ERC project (grant number 100117693), and by the Irish research Council (IRC) under grant number IRCLA- 2022-3788. Views and opinions expressed are however those of the author(s) only and do not necessarily reflect those of the European Union or the European Research Council Executive Agency. Neither the European Union nor the granting authority can be held responsible for them. SM acknowledges funding by the Royal Society through a Royal Society University Research Fellowship (URF-R1-221669) and the European Union through the FEED ERC project (grant number 101162711). AB acknowledges research support by the Irish Research Council under grant GOIPG/2022/1895. MRJ acknowledges support from the European Union's Horizon Europe Programme under the Marie Sklodowska-Curie grant agreement no. 101064124 and funding provided by the Institute of Physics Belgrade, through the grant by the Ministry of Science, Technological Development, and Innovations of the Republic of Serbia. PW acknowledges support from FONDECYT grant 3220399 and ANID -- Millennium Science Initiative Program -- Center Code NCN2024\_001. MB acknowledges funding from the Agence Nationale de la Recherche through the DDISK project (grant No. ANR-21-CE31-0015). AMH acknowledges support from the National Science Foundation under Grant No. AST-2307920. This work was also supported by the NKFIH NKKP grant ADVANCED 149943 and the NKFIH excellence grant TKP2021-NKTA-64. Project no.149943 has been implemented with the support provided by the Ministry of Culture and Innovation of Hungary from the National Research, Development and Innovation Fund, financed under the NKKP ADVANCED funding scheme. JPM acknowledges research support by the National Science and Technology Council of Taiwan under grant NSTC 112-2112-M-001-032-MY3. JM acknowledges funding from the Agence Nationale de la Recherche through the DDISK project (grant No. ANR-21-CE31-0015) and from the PNP (French National Planetology Program) through the EPOPEE project. SP acknowledges support from FONDECYT Regular 1231663 and ANID -- Millennium Science Initiative Program -- Center Code NCN2024\_001. A.A.S. is supported by the Heising-Simons Foundation through a 51 Pegasi b Fellowship. EC acknowledges support from NASA STScI grant HST-AR-16608.001-A and the Simons Foundation. JBL acknowledges the Smithsonian Institute for funding via a Submillimeter Array (SMA) Fellowship, and the North American ALMA Science Center (NAASC) for funding via an ALMA Ambassadorship. TDP is supported by a UKRI Stephen Hawking Fellowship and a Warwick Prize Fellowship, the latter made possible by a generous philanthropic donation. CdB acknowledges support from the Spanish Ministerio de Ciencia, Innovaci\'on y Universidades (MICIU) and the European Regional Development Fund (ERDF) under reference PID2023-153342NB-I00/10.13039/501100011033, from the Beatriz Galindo Senior Fellowship BG22/00166 funded by the MICIU, and the support from the Universidad de La Laguna (ULL) and the Consejer\'ia de Econom\'ia, Conocimiento y Empleo of the Gobierno de Canarias. This material is based upon work supported by the National Science Foundation Graduate Research Fellowship under Grant No. DGE 2140743. EM acknowledges support from the NASA CT Space Grant. Support for BZ was provided by The Brinson Foundation.
This paper makes use of the following ALMA data: ADS/JAO.ALMA\# 2022.1.00338.L, 2012.1.00142.S, 2012.1.00198.S, 2015.1.01260.S, 2016.1.00104.S, 2016.1.00195.S, 2016.1.00907.S, 2017.1.00167.S, 2017.1.00825.S, 2018.1.01222.S and 2019.1.00189.S. ALMA is a partnership of ESO (representing its member states), NSF (USA) and NINS (Japan), together with NRC (Canada), MOST and ASIAA (Taiwan), and KASI (Republic of Korea), in cooperation with the Republic of Chile. The Joint ALMA Observatory is operated by ESO, AUI/NRAO and NAOJ. The National Radio Astronomy Observatory is a facility of the National Science Foundation operated under cooperative agreement by Associated Universities, Inc. The project leading to this publication has received support from ORP, that is funded by the European Union’s Horizon 2020 research and innovation programme under grant agreement No 101004719 [ORP]. We are grateful for the help of the UK node of the European ARC in answering our questions and producing calibrated measurement sets. This research used the Canadian Advanced Network For Astronomy Research (CANFAR) operated in partnership by the Canadian Astronomy Data Centre and The Digital Research Alliance of Canada with support from the National Research Council of Canada the Canadian Space Agency, CANARIE and the Canadian Foundation for Innovation.

\end{acknowledgements}
 
\bibliographystyle{aa}
\bibliography{biblio}

\begin{appendix} 

\section{Keplerian mask creation}
\label{kep_mask}

We created the Keplerian masks used in the CLEANing process through the \verb|keplerian_mask| package \citep{rich_teague_2020_4321137}, with disc and stellar parameters as in Table \ref{tab:system_parameters}. As the CO lines at any given disc location, or pixel on the sky, are not infinitely thin in the spectral direction, we need to account for their intrinsic line widths, $dV$, in the mask-making process. We assume the intrinsic line width is set by Doppler broadening, with 

\begin{equation}
dV = 2 \sigma_v \sqrt{2\ln{2}} \text{ where }  \sigma_v = \sqrt{\frac{k_B T}{m_{\rm CO}}} ,     
\end{equation}

\noindent where $m_{\rm CO}$ is the mass of a CO molecule in kg. For simplicity we take the temperature $T$ of the gas to be the same as blackbody-like dust, leading to 

\begin{equation}
\label{bbtemp}
T = 278.3 {\rm K } \sqrt{\frac{L_{\star}^{0.5}}{r}}
\end{equation}

\noindent where $L_{\star}$ is the luminosity of the star in solar luminosities ($L_{\odot}$) and $r$ is disc radius in au. The assumed inner and outer radii of the CO discs for the mask are 

\begin{equation}
\label{rminmax}
    r_{\rm min/max} ["] = \frac{r_{\text{mid}} \pm 2\sigma_r}{d} \text{ where } \sigma_r = \frac{\Delta r }{2 \sqrt{2 \ln{2}}}
\end{equation}

\noindent and $d$ is the distance to the system in pc. $r_{\rm mid}$ is the radius to the centre of the disc and $\Delta r$ is the FWHM of the disc, both in au. As an initial guess, we take $r_{\rm mid}$ and $\Delta r$ for the dust grains' distribution in each system \citep[Table 3,][]{SebasARKSpaper}. 

\begin{figure}[h!]
    \centering
    \includegraphics[width=1\columnwidth]{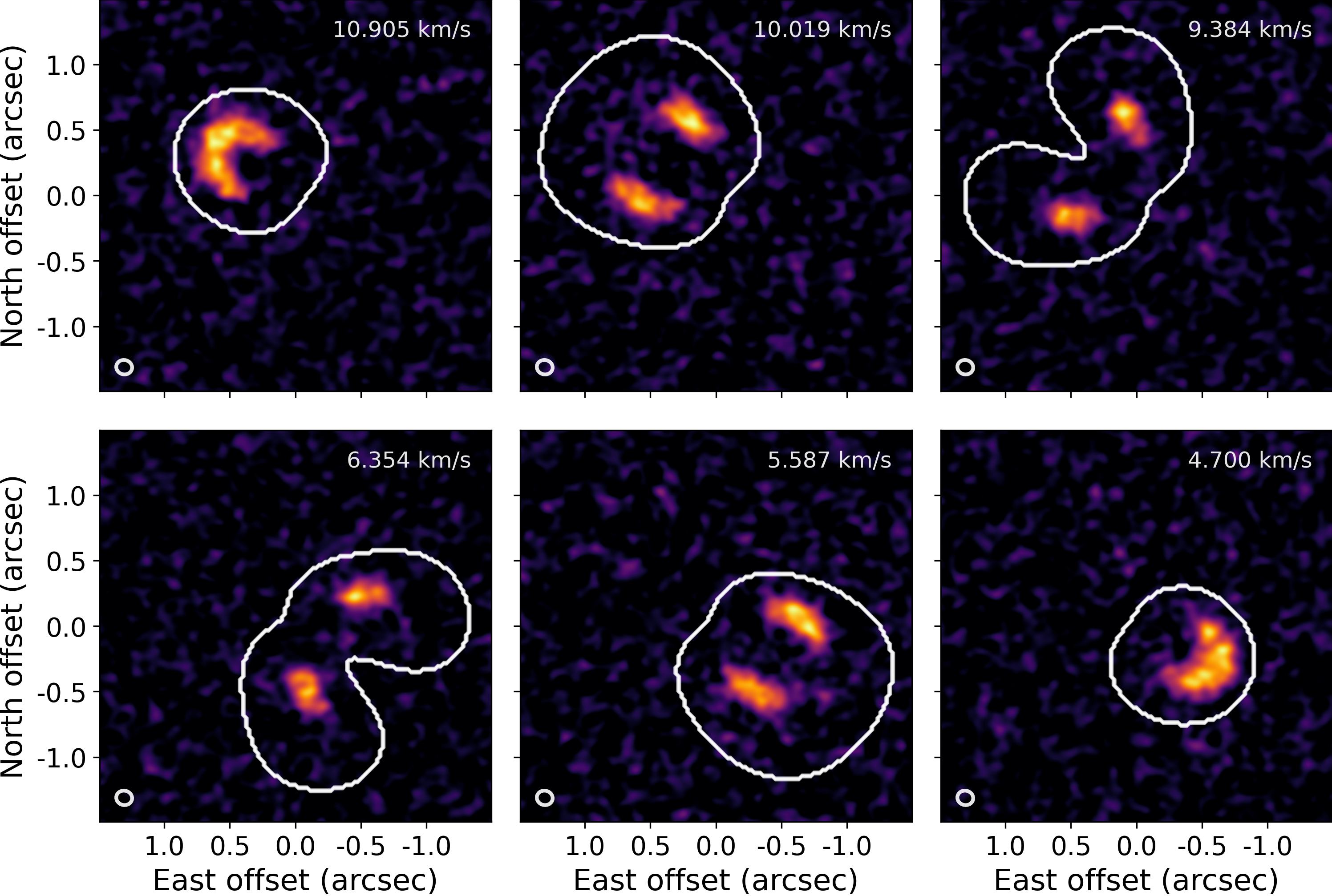}
    \caption{The Keplerian mask used during the imaging process shown here on channels from the HD~121617 $^{12}$CO image cube. The mask is created using stellar parameters (see Table \ref{tab:system_parameters}) assumes the gas is moving in Keplerian rotation around the star. The masks were created using the \texttt{keplerian\_mask} package \citep{rich_teague_2020_4321137}. The ellipse in the lower left corners is the synthesized beam of the observation.  } 
    \label{mask_gallery}
\end{figure}

The mask was then convolved with a Gaussian kernel of FWHM equal to \verb|nbeams| times the synthesised beam by using 
\verb|make_mask|'s \verb|nbeams| parameter. This ensures the mask covers more area than just the area covered by the Keplerian orbit. This parameter, effectively setting the area of the mask per channel, was adjusted after visually inspecting the cube to ensure all of the emission was included in the masks. Any inaccuracy in our assumptions to create the mask (e.g. gas radius, width, temperature, etc) are mitigated by the conservative choice of the \verb|nbeams| parameter smoothing in the final convolution step, which varied between 1.5 -- 10, depending on the system. The \verb|nbeams| parameter was particularly large for edge-on discs, for which it is not possible to assign a unique sky-projected velocity to each pixel, as each pixel probes a large range of orbital radii and azimuths. Examples of the mask surrounding the emission are shown in Fig. \ref{mask_gallery}, in which we show how the $^{12}$CO emission in HD~121617 was masked in a few example channels from the image cube.

\section{Beam information}
\begin{table*}[ht!]
\centering
\caption{Overview of the beam information for the ARKS $^{12}$CO and $^{13}$CO line observations. $\theta_{\rm maj} $ is the major axis and $\theta_{\rm min} $ the minor axis of the synthesised beam. The beam size listed is the mean size of the beam.}
\label{tab:beamsizes_gasrich}
\resizebox{\textwidth}{!}{
\begin{tabular}{c|cccc|cccc}
\multicolumn{1}{l|}{} & \multicolumn{4}{c|}{$^{12}$CO (13 m~s$^{-1}$ channels)} & \multicolumn{4}{c}{$^{13}$CO (443 m~s$^{-1}$ channels)}  \\ \hline
\multicolumn{1}{l|}{Gas-Bearing Discs} & \multirow{2}{*}{$\theta_{\rm maj} $ {[}"{]} $\times$ $\theta_{\rm min} ${[}"{]}} & {Beam Size}& Beam PA & RMS & \multirow{2}{*}{$\theta_{\rm maj} $ {[}"{]} $\times$ $\theta_{\rm min} ${[}"{]}} & {Beam Size} & Beam PA & RMS \\
\multicolumn{1}{l|}{} &  &  {[}au{]} & [$^\circ$] & {[}mJy/beam{]} &  & {[}au{]} & [$^\circ$] & {[}mJy/beam{]} \\ \hline \hline
HD~9672 & 0.16 $\times$ 0.13 &  8.3 & -87.5& 5.2 & 0.18 $\times$ 0.15 &  9.3 &89.6& 1.0 \\
HD~32297 & 0.09  $\times$ 0.08 &  11.3 & -56.5 &  3.8 & 0.09 $\times$  0.08 &  11.1 & -49.3 & 0.9 \\
HD~121617 & 0.12 $\times$ 0.11 &  13.5 &-76.5& 4.7 & 0.13 $\times$ 0.12 &  14.8 &-72.6 & 0.8 \\
HD~131488 & 0.08 $\times$  0.07 &  11.0 &-69.8 & 2.9 & 0.08  $\times$ 0.07 &  10.9 &-74.3 & 0.6  \\
HD~131835 & 0.13  $\times$  0.12 & 16.5 & 86.8 &3.9 & 0.14 $\times$ 0.13 &  18.3 &-89.8 & 0.9 \\ \hline 
\multicolumn{1}{l|}{Gas-Free Discs} & \multicolumn{4}{c|}{$^{12}$CO (423 m~s$^{-1}$ channels)} & \multicolumn{4}{c}{$^{13}$CO (443 m~s$^{-1}$ channels)} \\ \hline
HD~10647 & 0.58 $\times$  0.50 &  9.2 &-65.8 & 0.9 & 0.61 $\times$  0.52 &  9.7 &66.1 & 1.0 \\
HD~14055 & 0.37 $\times$  0.24 &   10.4&2.0 & 0.6 & 0.39 $\times$  0.26 &  11.0 &4.3 & 0.8 \\
HD~15115 & 0.20 $\times$  0.18 &  9.2 & 59.0 &0.4 & 0.22 $\times$  0.20 &  10.0 & 58.1 & 0.5\\
HD~15257 & 1.07 $\times$  0.82 &  47.3 &-39.6 & 1.5 & 1.15 $\times$  0.86 &  50.0& -39.8  & 1.6 \\
HD~61005 & 0.33 $\times$  0.22 &  9.9 &86.4& 1.3 & 0.40 $\times$  0.26 &  12.0 &86.6& 1.5 \\
HD~76582 & 0.80 $\times$  0.67 &  35.9 &61.8& 1.0 & 0.83 $\times$  0.70 &  37.5 &62.1& 1.1 \\
HD~84870 & 0.81 $\times$  0.57 &  62.2 &1.1& 0.9 & 0.85 $\times$  0.60 &  65.3&1.8 & 0.9 \\
HD~95086 & 0.49 $\times$  0.30 &  38.1 &17.2& 1.2 & 0.51 $\times$  0.41 &  39.8 &17.5& 1.6 \\
HD~109573 & 0.14 $\times$  0.12 &  9.5 &83.6& 1.3 & 0.16 $\times$  0.15 &  11.3 &88.9& 1.4\\
HD~145560 & 0.16 $\times$  0.13 &  17.2 &74.4& 0.6 & 0.17 $\times$  0.15 &  19.0 &72.3& 0.7 \\
HD~161868 & 0.80 $\times$  0.54 &  20.0 &-80.1& 1.1 & 0.83 $\times$  0.57 &  21.0 &-78.0& 1.2 \\
HD~170773 & 0.45 $\times$  0.39 &  15.6 &-66.9& 1.3 & 0.46 $\times$  0.41 &  16.1 &-67.7& 1.5 \\
TYC~93404371 & 0.73 $\times$  0.98 & 31.6 &-70.1& 1.3 & 0.76 $\times$  0.71 &  27.3 &-73.1& 1.7 \\  \hline \hline  
\end{tabular}
}
\end{table*}

\section{Disc parameters for ARKS
}

\begin{table*}[]
\centering
\caption{Disc parameters of the ARKS disc sample employed in this work. }
\label{tab:system_parameters}
\resizebox{0.95\textwidth}{!}{
\begin{tabular}{cccccccccc}
& i {[}$^{\circ}${]} & PA {[}$^{\circ}${]} & r {[}au{]} & $\Delta r$ {[}au{]} & M$_{*}$ [M$_{\odot}$] & d {[}pc{]} &  L$_{*}$ [L$_{\odot}$] &V$_{*}$ {[}m~s$^{-1}${]} & Ref. \\ \hline \hline 
HD~9672 &$78.7 \pm 0.2$	&$107.9 \pm 0.2$&	$135.2 \pm 1.9 $	&$158 \pm 3$ & 2.00	&	57.2	& 15 &$11906\pm 1 $	& 1\\ 
HD~32297  & $88.3 \pm 0.0$  & $47.5 \pm 0.0$   & $115.9 \pm 0.3$ & $37\pm 1$ & 1.57   & 130    & 7.0 & $21558^{+4}_{-3} $ & 1  \\   
HD~121617  & $44.1 \pm 0.6$    & $58.7 \pm0.7$   & $76.3\pm0.4$     & $18\pm1$      & 1.90  & 118   & 14.9 &$7851^{+3}_{-2} $  &  1 \\ 
HD~131488 & $85.0 \pm 0.1$  & $97.2 \pm 0.0$ & $89.7 \pm 0.1 $ & $13 \pm 1$  & 1.80   & 152   & 12 &$5860^{+7}_{-8}$  & 1 \\  
HD~131835  & $74.2\pm 0.2$  & $59.2\pm0.2$ & $77.1 \pm 0.5$ & $65\pm 2$   & 1.70  & 130  &  9.5  &$3621^{+6}_{-5}$  & 1 \\ \hline
HD~10647 & $77.8 \pm 0.1$	& $57.3 \pm 0.3$ &	$103.6 \pm 0.4$ &	$65 \pm 2$ 	& 1.12	& 17.3	&1.6 & $27600\pm100$  & 2 \\
HD~14055 & $80.7 \pm 0.2$	& $162.2 \pm 0.3$	& $169.4 \pm 2.9 $	&$162 \pm 9$ 	& 2.19	& 35.7	&29 & $5100\pm1700 $ &   2 \\
HD~15115 & $86.7 \pm 0.0$ &	$98.5 \pm 0.0$	&$80.7 \pm 0.3$ & $14\pm 1$ &	1.43 	& 48.8 &3.7& $6900\pm 300$ & 2 \\
HD~15257  & $59.4 \pm 4.9$ & $47.3 \pm 5.5$ & $184.2\pm 20.2$ & $261 \pm 41$  & 1.75  & 49.0  & 14 & $-24800\pm 2800$  &  3 \\ 
HD~61005& $85.9\pm 0.1 $& $70.4\pm 0.1$ & $72.4 \pm 0.2$ & $36 \pm 1$ & 0.95  & 36.5  &0.62  & $22300\pm 100$  & 2  \\ 
HD~76582 & $72.6 \pm 0.7 $ & $104.7 \pm 0.7 $ & $203.6 \pm 4.9$ & $176 \pm 9$ & 1.61 & 48.9  & 9.9 & $-1700 \pm 200 $ & 2 \\ 
HD~84870 & $47.1\pm 2.6$  & $-25.7\pm 4.1$ & $199.3\pm 7.8$ & $232\pm 15$  & 1.66 & 88.8  & 7.9  & $500\pm 200$ &  2  \\ 
HD~95086  & $31.6\pm 3.5$  & $100\pm 5.7$ & $197.0 \pm 4.6$ & $173\pm 8$  & 1.54  & 86.5   & 6.4 & $18000\pm 200$ & 2  \\ 
HD~109573 & $76.6 \pm 0.1$  & $26.5 \pm 0.0$  & $75.9 \pm 0.1$  & $7\pm 0$  & 2.14  & 70.8  & 25 & $10900 \pm 600$ & 2 \\ 
HD~145560  & $47.6\pm 0.6$ & $39.4\pm 0.9$ & $75.4\pm 0.5$  & $29 \pm 2$ & 1.35  & 121 &  3.3  & $ 2800\pm 500$  &  2  \\ 
HD~161868  & $66.1\pm 1.5$  & $57.6\pm 1.6$  & $121.2\pm 3.7$ & $134\pm 8$   & 2.11  & 29.7    & 24 & $-12300\pm 500$& 2 \\ 
HD~170773 & $32.9 \pm 1.9$  & $113.0 \pm 2.9$   & $191.9 \pm 2.9$ & $66\pm 4$  & 1.4 & 36.9 & 3.6  & $-17700 \pm 200$  &  2 \\
TYC~9340-437-1 & $18.0 \pm 8.0$  & $149.2\pm 20.2$  & $95.7\pm 2.5$ & $99 \pm 5$ & 0.75  & 36.7  &  0.19 &$7980\pm 20$ &  4 \\  \hline \hline
\end{tabular}
}
\caption*{\textbf{Notes:} The inclination, PA, disc radius ($r$) and disc width ($\Delta r$) (2$^{\rm nd}$, 3$^{\rm rd}$, 4$^{\rm th}$ and 5$^{\rm th}$ columns) are the best fit parameters derived from MCMC sampling carried out by \cite{SebasARKSpaper}. The listed values correspond to the median and the 16$^{\rm th}$ and 84$^{\rm th}$ percentiles to estimate the uncertainties. For sources with a double Gaussian (HD~107146, HD~92945 and HD~15115), $r$ represents the average between the two Gaussian centres and $\Delta r$ is the sum of the FWHM of the two Gaussians. Stellar masses in the 6$^{\rm th}$ column were estimated as described in Appendix A in \cite{SebasARKSpaper}. The 7$^{\rm th}$ column indicates the distance according to Gaia DR3 \citep{GDR3_2023}. The stellar luminosity (8$^{\rm th}$ column) were taken from \cite{matra2022_REASONS}.  The 9$^{\rm th}$ column indicates the stellar barycentric radial velocity. V$_{*}$ for gas-bearing systems were measured by fitting a Gaussian to the discs spectro-spatailly stacked spectra, while  V$_{*}$ for gas-free systems were taken from the literature; the corresponding reference is listed in the 10$^{\rm th}$ column. \newline \textbf{Velocity references:} 1. This work, 2. \cite{GDR3_2023}, 3. \cite{gontcharov2006pulkovo}, 4. \cite{zuniga2021search} }
\end{table*}

\section{Spectra and upper limits for non-detected discs}
\begin{figure*}[ht!]
    \centering
    \includegraphics[width=0.8\linewidth]{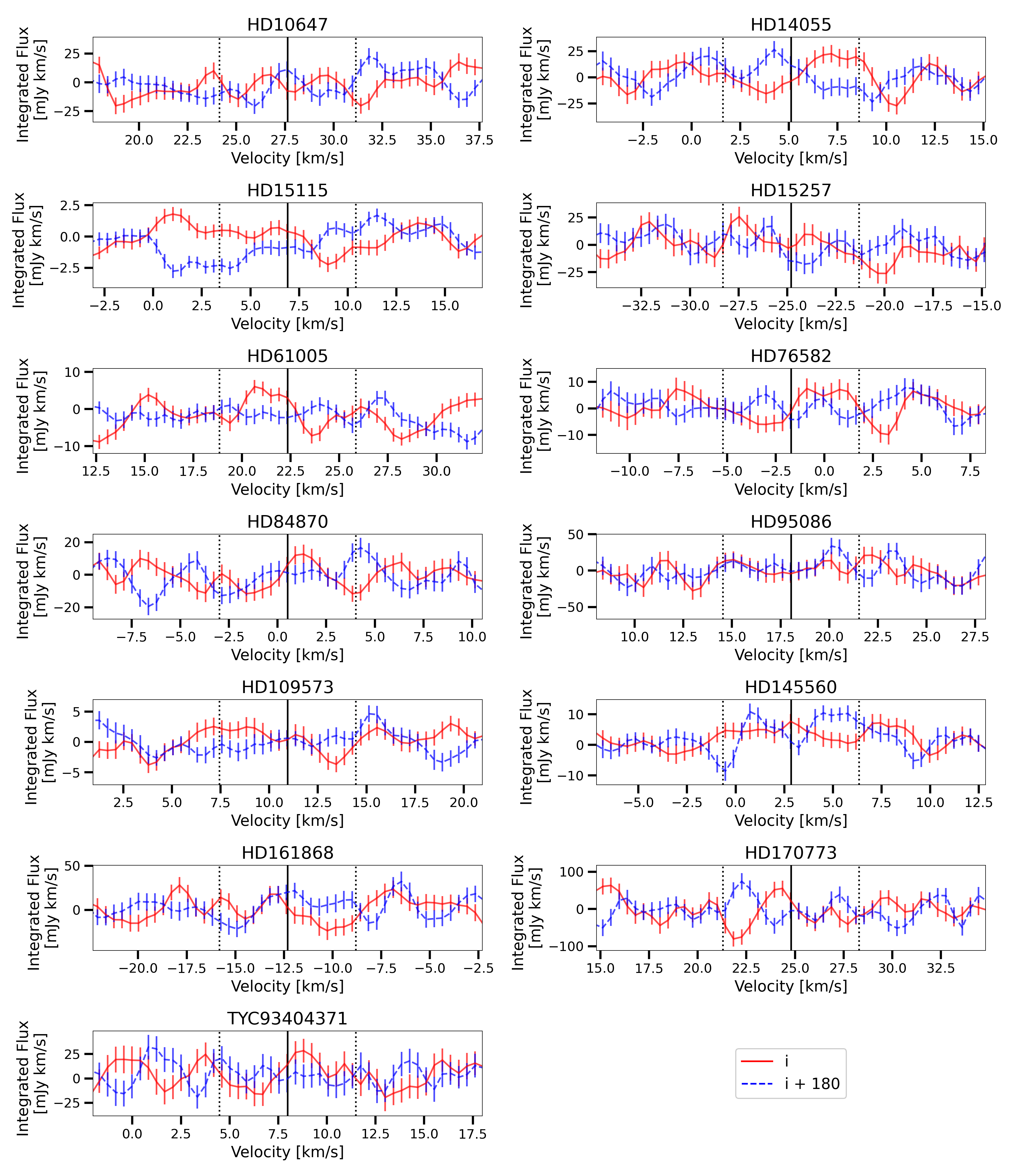}
    \caption{The spectro-spatially stacked spectra of the ARKS targets where no CO emission was detected. In each plot, we show both the spectrum shifted with the disc's inclination $i$ and also with $i + 180^\circ$, to eliminate the degeneracy between rotational direction and inclination. The black solid line indicates the stellar velocity, in the barycentric frame, which is where we would expect to see a peak if gas was detected in the system. The black dotted lines indicate the region of $V_* \pm 3.5$ km~s$^{-1}$ which was integrated over to calculate the line flux. }
    \label{gasfree_gallery}
\end{figure*}

\begin{table}[t!]
\centering
\caption{Upper limits on the integrated Line Flux of $^{12}$CO. }
\label{tab:gas_free_discs_line_flux}
\begin{tabular}{ccc}
\hline
Gas-free Discs & \begin{tabular}[c]{@{}c@{}}3 $\sigma$ upper limit on\\  Integrated line flux of $^{12}$CO \\
{[}mJy~km~s$^{-1}${]}\end{tabular} & Reference\\ 
\hline\hline
HD~10647       & 84  &1 \\
HD~14055       & 83  &1\\
HD~15115       & 20  &1\\
HD~15257       & 85  &1\\
HD~61005       & 59  &1\\
HD~76582       & 49  &1\\
HD~84870       & 44  &1\\
HD~95086       & 100 &1\\
HD~109573      & 20  &1\\
HD~145560      & 67  &1\\
HD~161868      & 230 &1\\
HD~170773      & 130 &1\\
TYC~93404371   & 160 &1\\ \hline
HD~92945&69& 2 \\
HD~107146&74 & 3 \\
HD~197481&70 $^a$ & 4 \\
HD~206893& 40 $^a$ & 5\\
HD~218396& - $^b$ & 6 \\
 \hline
\end{tabular}
\caption*{\textbf{Notes. }Archival upper limits on the $^{12}$CO (J=3-2) line are included for the ARKS archival targets. $^{(a)}$ denotes $^{12}$CO (J=2-1) is reported. $^{(b)}$ No upper limit is available for HD~218396, as discussed in Appendix B of \cite{faramaz2021detailed}. \textbf{References:} 1. This work, 2. \cite{marino2019gap}, 3. \cite{marino2018gap}, 4. \cite{daley2019mass}, 5. \cite{nederlander2021resolving} 6. \cite{faramaz2021detailed}}
\end{table}

\section{Archival $^{12}$CO integrated line fluxes}
\begin{table*}[]
\caption{Archival $^{12}$CO integrated line fluxes plotted in Figs. \ref{scaledfluxage} and \ref{scaledfluxfraclum}. $^{12}$CO J=2-1 integrated line fluxes were multiplied by the average $^{12}$CO J=3-2/J=2-1 ratio (1.67) before being plotted along side the $^{12}$CO J=3-2 integrated line fluxes.}
\label{tab:archival_linefluxes}
\resizebox{\textwidth}{!}{
\begin{tabular}{c|cc|cc|ccc}
\textbf{} & \textbf{Age } & \textbf{Ref.} & \textbf{Fractional Luminosity} & \textbf{Ref.} & \textbf{$^{12}$CO Line} & \textbf{$^{12}$CO Integrated Flux} & \textbf{Ref.} \\
& \textbf{[Myr]} &  &  &  &  & \textbf{[mJy~km~s$^{-1}$]} &  \\ \hline \hline 
HD~21997 & $42^{+6}_{-4}$ & 1 & $(5.8\pm 0.1)\times10^{-4}$ & 1 & 3-2 & $2513\pm260$ & 1 \\
HD~36546 & $6\pm4$ & 2 & $(4.7\pm 0.1)\times10^{-3}$ & 1 & 2-1 & $2670\pm40$ & 2 \\
HD~38206 & $42^{+6}_{-4}$ & 1 & $(1.16\pm 0.04)\times10^{-4}$ & 1 & 2-1 & $<18$ & 3 \\
HD~39060 & $23\pm3$ & 1 & $(2.5\pm 0.1)\times10^{-3}$ & 1 & 3-2 & $5807\pm606$ & 4 \\
HD~48370 & $36\pm 8$ & 4 & $(5.6\pm 0.2)\times 10^{-4}$ & 2 & 2-1 & $<338$ & 1 \\
HD~98363 & $15\pm3$ & 6 & $9.20\times 10^{-4}$ & 3 & 2-1 & $<36$ & 7 \\
HD~106906 & $15\pm3$ & 1 & $(1.2\pm0.03)\times10^{-3}$ & 1 & 2-1 & $<43$ & 8 \\
HD~109085 & $1500^{+1000}_{-500}$ & 7 & $(1.7\pm0.4)\times10^{-4}$ & 1 & 3-2 & $38\pm9$ & 9 \\
H~109832 & $15$ & 8 & $7.6\times10^{-4}$ & 4 & 2-1 & $<33$ & 7 \\
HD~110058 & $15\pm3$ & 1 & $(9^{±22}_{-9})\times10^{-4}$ & 1 & 3-2 & $191\pm17$ & 1 \\
HD~111520 & $15\pm3$ & 1 & $(2.30\pm0.06)\times10^{-3}$ & 1 & 2-1 & $<43$ & 8 \\
HD~113766 & $15\pm3$ & 1 & $(2.2^{+0.7}_{-0.3})\times10^{-2}$ & 1 & 2-1 & $<45$ & 8 \\
HD~114082 & $15\pm3$ & 1 & $(3.55^{+0.07}_{0.08})\times10^{-3}$ & 1 & 2-1 & $<45$ & 8 \\
HD~115600 & $15\pm3$ & 1 & $(1.95^{+0.05}_{-0.04})\times10^{-3}$ & 1 & 2-1 & $<43$ & 8 \\
HD~117214 & $15\pm3$ & 1 & $(2.66^{+0.07}_{-0.06})\times10^{-3}$ & 1 & 2-1 & $<39$ & 8 \\
HD~121191 & $16\pm2$ & 1 & $(3.2^{+0.6}_{-0.3})\times10^{-3}$ & 1 & 2-1 & $209\pm19$ & 10 \\
HD~129590 & $16\pm2$ & 1 & $(5.5^{+0.2}_{-0.1})\times10^{-3}$ & 1 & 2-1 & $<54 $& 8 \\
HD~138813 & $10\pm3$ & 1 & $(7.7\pm0.2)\times10^{-4}$ & 1 & 2-1 & $1406\pm78$ & 8 \\
HD~141569 & 5 & 9 & $8 \times 10^{-3}$ & 5 & 3-2 & $15800\pm600$ & 11 \\
HD~142315 & $10\pm3$ & 1 & $(6.6\pm0.1)\times10^{-4}$ & 1 & 2-1 & $<54$ & 8 \\
HD~146181 & $16\pm2$ & 1 & $(1.9\pm0.2)\times10^{-3}$ & 1 & 2-1 & $<36$ & 8 \\
HD~146897 & $10\pm3$ & 1 & $(8.0^{+0.4}_{-0.3})\times10^{-3}$ & 1 & 2-1 & $60\pm15$ & 8 \\
HD~147137 & $10\pm3$ & 1 & $(4.4^{+0.1}_{-0.2})\times10^{-4}$ & 1 & 2-1 & $<43$ & 8 \\
HD~156623 & $16\pm7$ & 10 & $3.778\times 10^{-3}$ & 6 & 2-1 & $1183\pm37$ & 8 \\
HD~172555 & $23\pm3$ & 11 & $7.2 \times 10^{-4} $ & 7 & 2-1 & $170 \pm 30$ & 12 \\
HD~181327 & $23\pm3$ & 1 & $(2.71\pm0.05)\times10^{-3}$ & 1 & 3-2 & $<199$ & 1 \\
NO Lup & 1-3 & 13 & $(3.4 \pm 0.01)\times 10^{-3}$ & 8 & 3-2 & $290 \pm70$ & 14 \\
TWA7 & $10\pm3$ & 1 & $(2.10\pm 0.08)\times10^{-3}$ & 1 & 3-2 & $91\pm20$ & 15 \\
Fomalhaut & $440\pm40$ & 14 & $(1.41\pm0.02)\times10^{-5}$ & 1 & 2-1 & $68\pm16$ & 16 \\ \hline \hline 
\end{tabular}
}
\caption*{\textbf{Age References:} 1. \cite{matra2022_REASONS}, 2. \cite{currie2017subaru}, 3. \cite{torres2008young}, 4. \cite{gratton2024stellar} 5. \cite{booth2021age}, 6. \cite{melis2013copious},  7. \cite{pearce2022planet} and references therein, 8. \cite{pecaut2016star}, 9. \cite{flaherty2016resolved}, 10. \cite{mellon2019discovery}, 11. \cite{schneiderman2021carbon}, 12. \cite{kennedy2018alma}. 13. \cite{lovell2021rapid}, 14. \cite{mamajek2012age}.  \textbf{Fractional Luminosity References:} 1. \cite{matra2022_REASONS}, 2. \cite{moor2016new}, 3. \cite{chen2014spitzer}, 4. \cite{ballering2013trend}, 5. \cite{sylvester1996optical},  6. \cite{mcdonald2012fundamental}, 7. \cite{wyatt2007transience}, 8. \cite{lovell2021alma}. \textbf{Flux References:} 1. \cite{cataldi2023primordial}, 2. \cite{rebollido2022search}, 3. \cite{booth2021resolving}, 4. \cite{matra2017exocometary_betapic}, 5. \cite{booth2019deep}, 6.  \cite{booth2021age} 7. \cite{moor2017}, 8. \cite{lieman2016debris}, 9. \cite{marino2016alma}, 10. \cite{kral2020survey}, 11. \cite{flaherty2016resolved}, 12. \cite{schneiderman2021carbon}, 13. \cite{kennedy2018alma}, 14. \cite{lovell2021rapid},  15. \cite{matra2019kuiper}, 16. \cite{matra2017detection_fomalhaut}}
\end{table*}

\section{Discussion on individual discs}
\label{App:Indiv_discs}
\subsection{HD~9672 (49 Ceti)}
\label{disc:HD9672}
HD~9672 is the only system for which the peaks of the $^{12}$CO and $^{13}$CO are not co-located (Fig. \ref{RadProf_gallery}). The $^{12}$CO is centrally peaked at $\sim 33$ au, the $^{13}$CO peaks much farther out in the system, at $\sim 73$ au, while the dust continuum peaks at $135.2 \pm 1.9$ au \citep{SebasARKSpaper}. Both $^{12}$CO and (marginally) $^{13}$CO are detected all the way to the central resolution element of the observations (the velocity-based inner radius of the $^{12}$CO is $\sim 3$ au) and out as far as $\sim 200$ au. The difference in peak locations between $^{12}$CO and $^{13}$CO could be an effect of the $^{12}$CO being optically thick, while the $^{13}$CO is optically thin. Optically thick $^{12}$CO would trace the temperature radial distribution (hotter closer to the star), while the optically thin $^{13}$CO traces more closely the surface density of the gas. HD~9672's $^{12}$CO/$^{13}$CO ratio displays (at least marginally) what we would expect from optically thick $^{12}$CO co-located with optically thin $^{13}$CO, with the ratio at its lowest at the peak location of the $^{13}$CO and increasing away from the peak. 

In its 1D spectral profiles (Fig. \ref{RadProf_gallery}, left column), we see the $^{12}$CO shows a very steep, V-shaped profile interior to its two spectral peaks, whereas $^{13}$CO shows a shallower, U-shaped profile. This is further evidence towards $^{12}$CO being optically thick with optically thinner $^{13}$CO \citep{horne1986emission}, making HD~9672 unique in the ARKS sample. 3D radiative transfer modelling and an analysis of the local line profiles is needed to determine the optical depth and more accurately determine the gas mass in this system, similar to what is presented in \citet{Aoife_arks} and \citet{Marija_arks} for HD~121617 and HD~131835.

\cite{hughes2017radial} had previously measured the integrated flux of the $^{12}$CO (3-2) in HD~9672 to be $6.0 \pm 0.1$ Jy km s$^{-1}$ (though the 10\% flux calibration uncertainty is not included in this error bar), measured from the flux enclosed by the 3$\sigma$ contours on their integrated-intensity map. With the improved spectral and spatial resolution, we have measured a higher value of the integrated flux of $^{12}$CO in the disc to be $8.9 \pm 0.9$ Jy km s$^{-1}$. The increased value is likely due to the $3~\sigma$ contour measurement by \cite{hughes2017radial} missing some extended, low-level emission of the outer disc. They calculated an expected outer radius of the CO emission to be $\sim 220$ au. With the improved spectral and spatial resolution offered by ARKS, we can reduce this outer boundary: we find an outer radius of $\sim 200$ au (Fig. \ref{RadProf_gallery}). We confirm the existence of the asymmetry between the south-east and north-west limbs first reported in \cite{hughes2017radial} in which the CO emission was twisted relative to the disc major axis. We observe that the $^{12}$CO emission in the north west is brighter than the emission in the south east of the disc; in addition, on the NW side, the brightness peak is south of the disc major axis, which is in the same direction as the twist previously reported for $^{12}$CO \citep{hughes2017radial} and now observed in the ARKS continuum observations \citep{Josh_arks}. This is clear for $^{12}$CO from its integrated-intensity map, but not clearly seen for the fainter $^{13}$CO (Fig. \ref{12CO_gallery}), therefore it remains unclear for CO whether this reflects a surface density asymmetry or a temperature asymmetry.

\subsection{HD~32297}
\label{disc:HD32297}
HD~32297 is a close to edge-on ($i = 88.33^\circ$), vertically thin disc when observed in the continuum \citep{Brianna_arks}, though broader when observed through its $^{12}$CO and $^{13}$CO emission and in scattered light observations \citep{Julien_arks}. From the on-sky (non-deprojected) radial profile, CO radially extends as far out as $\sim 200$ au, peaking around $\sim 75 $ au and potentially all the way to the inner regions of the planetary system (Fig. \ref{RadProf_gallery}), though the velocity-based upper limit on the inner radius of the $^{12}$CO is $\sim 68$ au. This outer radius is consistent with what was observed with low resolution observations; \cite{moor2019new} also found an outer radius of $\sim 200$ au. The mm dust, however, peaks at $105.2^{+0.7}_{-0.5}$ au, in a much narrower belt \citep[$\Delta R = 14 \pm 0.6$ au,][]{Yinuo_arks}. 

We observe a asymmetry between the NE and SW sides of the disc in the gas, with the peak emission being brighter and more extended on the SW side of the disc, as seen in the moment maps (Fig. \ref{12CO_gallery}), and as was seen for CI in Fig. 1 of \citet{cataldi2020surprisingly}. However, the radial profile shows that the NE side of the disc is brighter in the outer ($>100$ au) regions of the disc (Fig. \ref{RadProf_gallery}). This disc is observed to be asymmetric in other wavelengths; \cite{Josh_arks} determines the system to host an offset asymmetry while \cite{Julien_arks} shows the disc is more extended in small grains in the NE (as is the case in the radial profiles in Fig. \ref{RadProf_gallery}). All combined, this suggests that the disc is eccentric, with its orbital apocentre in the same direction as the asymmetries presented here. 

HD~32297's $^{12}$CO/$^{13}$CO ratio does not change with radius and is constant at $\sim 2$. While this ratio is between the non-deprojected, on-sky radial profiles, it shows that both $^{12}$CO and $^{13}$ CO are likely both optically thick or both optically thin, as discussed above. 

\subsection{HD~121617}
\label{disc:HD121617}
The bulk of the $^{12}$CO emission in HD~121617 is confined to a ring, peaking at $ \sim 73 $ au and $\sim 48$ au wide (FWHM from fitting a Gaussian to the radial profile) in $^{12}$CO (Figs. \ref{12CO_gallery}, \ref{RadProf_gallery}). However, interior to this ring the emission does not go down to zero, instead reaching a minimum at $\sim 34$ au and rising again interior to that all the way to the central resolution element. The velocity-based upper limit on the inner radius of the $^{12}$CO is $\sim 28$ au. Similarly, outside the main ring, $^{12}$CO emission is detected out to $\sim 160$ au in a long, faint tail of emission. This is in contrast to the mm-sized dust grains, which peak at a slightly larger radius, $r= 75.3^{+0.4}_{-0.3}$ au, than the $^{12}$CO intensity profile, but are much more narrowly distributed \citep[FWHM = $14.1 \pm 1.2$ au for the mm grains vs $\sim 48$ au for the $^{12}$CO,][]{Yinuo_arks}. 

The $^{12}$CO/$^{13}$CO ratio is $\sim 2$ at the peak location of the $^{12}$CO and is largely consistent with being flat throughout the planetary system where both $^{12}$CO and $^{13}$CO are significantly detected. Between 60 au and 90 au, the $^{12}$CO/$^{13}$CO ratio increases slightly to $\sim 3$, which is likely caused by the $^{13}$CO becoming less optically thick with radius. We observe two large peaks in the $^{12}$CO/$^{13}$CO ratio for this system, located at $\sim 35$ au and $\sim 105$ au, in Fig. \ref{12_13_Ratio_gallery} but we note that these correspond to locations where the $^{13}$CO becomes only marginally detected and should not be interpreted as a change in the optical depth of the isotopologues.

\citet{Aoife_arks} carried out detailed radiative transfer modelling of this disc and found that the $^{12}$CO and $^{13}$CO emission are best explained by a narrow disc of gas, that resembles the thin dust disc, that is optically thick at all radii. They also analyse the intrinsic line profiles of the $^{12}$CO and $^{13}$CO and find them to be Gaussian, which was previously believed to indicate optically thin gas. However, they find their optically thick model can reproduce broad Gaussian line profiles. Their $^{13}$CO RADMC-3D best-fit model reproduces the observations with $T_{\rm peak} = 38$ K, and an optically thick $^{13}$CO mass of $M_{^{13}\rm CO} = 2 \times 10^{-3}$ M$_\oplus$. This was then scaled by the ISM $^{12}$CO/$^{13}$CO ratio of 77 to yield a total CO mass of $M_{\rm CO} = 1.48 \times 10^{-1}$ M$_\oplus$. We note here that the mass calculated in this work, presented in Table \ref{tab:linefluxesofgas}, assumed an optically thin $^{13}$CO, with $\tau_{^{13}\rm CO} = 1$, whereas $\tau_{^{13}\rm CO} = 26$ in the best-fit model of \cite{Aoife_arks}, which explains the discrepancy in the reported mass values. Their $T_{\rm peak}$ is comparable with the $T_{\rm peak}$ measured for this system from the $^{12}$CO peak-intensity/brightness temperature map. 

The continuum observations for this disc show a strong asymmetry with a brighter arc of emission at the SW ansa of the main ring \citep{Josh_arks, hd121617_arks}. However, this asymmetry is not seen in the CO emission (Fig. \ref{12CO_gallery}). By fitting a parametric radiative transfer model to the mm-sized dust grains, \cite{hd121617_arks} constrained the arc morphology and found the arc to be radially narrow and only marginally resolved with a FWHM between $\sim 1 - 5$ au. They examined the gas kinematics of the ring and found that the azimuthal velocity of the gas decreases radially more steeply than Keplerian, which could be explained by a strong pressure gradient. \cite{hd121617_arks} concluded that the velocity deviations are consistent with a narrow ring of gas, and that this asymmetry could be caused by a vortex in the gas \citep{Phillip_arks} or by planet-disc interactions, (Pearce et al, in prep.). 

In addition, to investigate if the CO disc displays any shift in position due to the disc being eccentric \citep[as found in scattered light emission, ][]{perrot2023morphology, Julien_arks}, \cite{hd121617_arks} examined the differences between the radial profiles of the north east and south west sides of the disc and found there is a significant signal when examining these differences, indicating there could be an eccentricity in the CO emission. However, this finding is sensitive to the stellar location assumed and due to an uncertainty of ${\lesssim}10$ mas on the stellar location \citep{SebasARKSpaper}, this eccentricity in the CO emission cannot be confirmed. 

The asymmetry only being detected in the continuum observations and not in the gas emissions could be due to a) the high optical depth of CO emission, which could hide an overdensity; or b) the gas distribution is approximately axisymmetric. The latter is consistent with the study of \cite{Phillip_arks}, who investigate the structure around HD~121617 via hydrodynamical simulations. Here, the narrow gas ring becomes unstable and generates a subtle azimuthal asymmetry in the form of a vortex. The size-dependent gas drag influences both the radial and azimuthal confinement of the dust particles. However, \cite{Phillip_arks} finds that the resultant dust structure is only consistent with the ALMA continuum and VLT/SPHERE scattered light image, if there is substantial unseen molecular hydrogen (i.e. if the total gas mass exceeds the measured CO mass by orders of magnitude). This assumption only makes sense in the primordial gas scenario. For the secondary gas scenario, the large dust and gas might be dynamically decoupled and hence evolve independently after being released through collisions.

\subsection{HD~131488}
\label{disc:HD131488}
HD~131488 is the system in which the CO peaks closest to the star, at $\sim15$ au in the on-sky (non-deprojected) radial profile (Fig. \ref{RadProf_gallery}). The CO extends outwards to $160$ au and potentially extends inwards as far as the innermost resolution element (the highest velocity and on-sky along-midplane distance from the star at which a $3~\sigma$ detection is achieved anywhere in the system corresponds to $r_{\rm min} \sim 6$ au). 

The $^{12}$CO radial profile (Fig. \ref{RadProf_gallery}) is similar to what might be expected in a primordial protoplanetary-like disc, and could point to a primordial origin if both $^{12}$CO and $^{13}$CO are optically thick (Sect. \ref{discussion_isotopologue_ratios}). In the radial profiles for HD~131488, we see that the CO extends to small radii ($r_{\rm min}<6$ au), suggesting that gas might be accreting onto the star. However, \citet{melis2013copious} presented the optical spectrum of HD~131488 and reported no evidence of accretion at a detectable level, as no noticeable emission lines tracing accretion nor any obvious filling of the Balmer absorption lines were present in their spectra.

For CO to be of second-generation origin, on the other hand, it needs to have had enough time to viscously spread inwards, from the thin dust disc, which peaks much farther out in the planetary system \citep[$r_{\rm mm} = 90.84^{+0.18}_{-0.08}$ au,][]{Yinuo_arks}. To check whether this viscous spreading could have propagated/evolved as far as the star, we calculate the viscous timescale, $t_\nu(r)$, at the location of the dust outer ring: 

\begin{equation}
    t_\nu(r) = \frac{r^2}{v_\nu} = \frac{\mu \sqrt{GM_* r}}{\alpha k_B T} \hspace{1cm},
\end{equation}

where $v_\nu = \alpha c_s H$ is the viscosity parametrized by the $\alpha$ parameter, $\alpha$ is the viscosity of the gas, $c_s$ is the sound speed and $H = \frac{c_s}{\Omega}$ is the local disc scale height \citep{shakura1973black, kral2016self}. We calculate $c_s$ by setting the molecular weight $\mu = 28$ and $T = 20$ K, measured from the $^{12}$CO peak intensity/brightness temperature map at $90$ au. We adopt the values of $\alpha$ that \cite{kral2019imaging} found can be used to explain the high CO mass in this disc due to its high CO input rate, $10^{-2} < \alpha <1$. This yields viscous timescales for the $^{12}$CO between $0.28$ Myr and $28.23$ Myr. This system is 16 Myr old \citep{matra2022_REASONS}, so this simple calculation suggests that the gas may have had sufficient time to spread in as far as the star. A caveat here is that while this is the entirety of the gas, this applies to CO only as long as its photodissociation timescale is longer than this viscous timescale, which depends on its column density perpendicular to the midplane (leading to self-shielding), and the presence of CI shielding, whose derivations would require more detailed modelling than presented here \citep{marino2020population, marino2022vertical}.

Thus, if the viscosity of the gas were sufficiently high for the viscous timescale to be shorter than the system age, to not be accreting onto the star \citep{melis2013copious}, the gas disc must be truncated and have an inner radius smaller than $r_{min} < 6 $ au. A potential explanation for this truncation could be the presence of a planet in the inner regions of the system. \citet{marino2020population} reported that a planet orbiting interior to a debris disc could stop the inward flow of gas released by the disc's icy planetesimals. To calculate its mass, we must consider the viscous limit of the gaseous disc, which requires the gap be opened faster than the viscous gas can refill it \citep{LinPapaloizou1993, duffell2013gap}.  We can also estimate the mass of a planet sculpting a gap in a gaseous disc by measuring the width of the gap and using the planet's Hill sphere \citep{bergez2024observing}. We set Eq. 6 in \cite{duffell2013gap} equal to Eq. 6 in \citet{bergez2024observing} to calculate the required mass of a planet to open a gap in a viscous gas and to find where in the planetary system the planet is likely to be located. For this, we measure the temperature at the highest velocity and on-sky along-midplane distance from the star at which a $3~\sigma$ detection is achieved anywhere in the system, corresponding to $r_{\rm min} \sim 6$ au, by converting the $3~\sigma$ pixel's intensity to brightness temperature using Planck's law. We then extrapolate this temperature throughout the inner regions of the system by assuming a power law $T(r) \propto r^{-0.5}$. We assume that the gas is dominated by CO ($\mu = 28$), and that the viscosity of the gas is the minimum viscosity that would have allowed the gas to viscously spread throughout the system in the star's lifetime, $\alpha = 0.017$. We find that a $7 M_\oplus$ planet located at 5.8~au is capable of opening a gap whose outer edge corresponds to the inner radius of the gas disc, 6~au. Thus, if such a planet exists in the system, we can infer its orbital radius is $\lesssim 5.8$~au and its mass $\gtrsim 7  M_\oplus$, though we note that the viscous limit of the gaseous disc will scale linearly with the viscosity of the gaseous disc, so these planet parameters should be interpreted as a lower limit on the planet mass and an upper limit on the planets location. This disc also has two dust rings with a gap in-between them at $\sim 69$ au, which \cite{Yinuo_arks} estimates a planet with a mass of $1.3 M_{\rm Jup}$ and orbital distance of $69.4$~au could have sculpted. Thus, there are indications for multiple planets in this system. 

As reported in \citet{Yinuo_arks}, the dust disc in this system peaks at $90.84^{+0.18}_{-0.08}$ au. An entirely different possibility to explain the strong difference in peak location between CO and dust is that at least some of the gas is produced in situ in the inner region of the system, where planets may be forming. \cite{melis2010age} reported significant amounts of warm  dust in the system, located at a radius of $0.6$ au under the assumption that the dust grains temperature is given by their black body radiation. Departures from this assumption, however, may shift the radius further out, that is, closer to the observed CO peak location. Therefore, CO gas may be connected to the presence of warm dust and potentially giant impact(s) expected in the terrestrial region for a system of this age, and as observed more clearly for similarly aged A star HD~172555 \citep{schneiderman2021carbon}. However, this might not be sufficient to explain why gas extends as far out as $>160$ au in this system, which likely requires an additional source of gas at such large distances. 

\subsection{HD~131835}
\label{disc:HD131835}
HD~131835 is a moderately inclined disc, at $i = 74.2^\circ \pm 0.2^\circ$ \citep{SebasARKSpaper}, which is now resolved over several resolution elements in the integrated-intensity and the peak-intensity maps (Fig. \ref{12CO_gallery}). The $^{12}$CO peaks at $\sim 63$ au while the mm-sized dust peaks at $ 68.6 \pm 0.4$ au \citep[Fig. \ref{RadProf_gallery}, ][]{Yinuo_arks}. We observe the CO out in the planetary system as far as $\sim 220$ au and calculate the velocity-based upper limit on the inner radius to be $\sim 21$ au. As in the other gas-bearing discs, the gas distribution is much broader than that of the mm-sized dust, which again could indicate a secondary origin for the CO which is now viscously spreading throughout the planetary system. This is consistent with spatially unresolved data analysed by \cite{hales2019modeling} of this disc, from which they argued that the partial co-location of the gas and dust supports a cometary origin for the gas. 

The $^{12}$CO radial profile (Fig. \ref{RadProf_gallery}) shows a marginal dip at $\sim 30$ au. This dip is not seen clearly in the images of the $^{12}$CO (Fig. \ref{12CO_gallery}), so higher resolution observations are needed to confirm its existence. In the $^{12}$CO/$^{13}$CO ratio radial profile, we find the ratio is constant with radius until $\sim 80$ au, which as previously discussed in Sect. \ref{discussion_isotopologue_ratios}, could indicate both isotopologues being optically thin or both optically thick. However, exterior to $\sim 80$ au, we see a marginal increase in the ratio, which could indicate the $^{13}$CO is becoming less optically thick at these large radii and is co-located with optically thick $^{12}$CO, as discussed in Sect. \ref{discussion_isotopologue_ratios}, although within the $3\sigma$ uncertainties, the $^{12}$CO/$^{13}$CO ratio remains consistent with being radially flat. 

\cite{moor2015discovery} previously measured these using APEX and found the integrated line flux of $^{12}$CO J=3-2 to be $2.74 \pm 0.55$ Jy~km~s$^{-1}$, which is consistent with our ARKS measurement at the $3 ~\sigma$ level (Table \ref{tab:linefluxesofgas}).

There is a large offset of $\sim 35 $ au between the peak emission from mm-sized dust grains and the peak of scattered light from micron-sized dust grains in this system \citep{Julien_arks}. Extensive modelling has been carried out to try to explain these differences, through two possible scenarios \citep{Marija_arks}. In the first scenario, the disc contains two planetesimal belts which produce two dusty rings with very different size distributions, such that they appear very differently in mm emission and in scattered light. This scenario matches observations to a large degree, albeit the dust in the two planetesimal belts needs to have extremely different material strength, or alternatively the belts need to have some combination of different dynamical excitation and different composition. It is not clear if these conditions can be fulfilled realistically.

In the second scenario, the peak of mm dust emission is tracing a planetesimal belt, and the outer ring is formed by smaller dust grains that have drifted outwards due to gas drag. Relatively simple dynamical models show that, while the location of the outer ring and the relative brightness of the rings in scattered light can be produced for reasonable assumptions, the outer ring would be much fainter in thermal emission than what is observed. This disagreement with observations could be resolved with a more detailed model that includes dust growth, or if there are planetesimals producing dust at large separations as well (but which do not need to be any different that the ones closer in, in this scenario). In any case, for this scenario to work, the gas mass cannot be as low as the optically thin CO mass found in this work, or else the gas drag would act too slowly on even the smallest grains.

\end{appendix}
\end{document}